\documentclass[twocolumn]{aastex631} % Choose the style appropriate for your submission
\usepackage[utf8]{inputenc}
\usepackage{natbib}
\usepackage{graphicx}
\usepackage{hyperref}
\usepackage{float}
\usepackage{rotating}
\usepackage{tablefootnote}
\usepackage{lineno}
\usepackage{longtable}

\usepackage{xspace}

\renewcommand\nodata{ ~$\cdots$~ }%
\newcommand{\hide}[1]{}

\newcommand{\y}[1]{\ensuremath{y_{#1}}\xspace}

\renewcommand{\micron}{\mbox{$\mu$m}}%

\newcommand{\kms}{\ensuremath{\,{\rm km\,s^{-1}}}\xspace}

\newcommand{\m}{\ensuremath{\,{\rm m}}\xspace}

\newcommand{\K}{\ensuremath{\,{\rm K}}\xspace}

\newcommand{\myr}{\ensuremath{\,{\rm Myr}}\xspace}

\newcommand{\h}[1]{\ensuremath{^{#1}{\rm H}}\xspace}

\newcommand{\hii}{{\rm H\,{\footnotesize II}}\xspace}

\newcommand{\cii}{{\rm [C\,{\footnotesize II}]}\xspace}
\newcommand{\oi}{{\rm [O\,{\footnotesize I}]}\xspace}
\shorttitle{Expansion Signatures in 35 \hii\ Regions traced by SOFIA \cii\ Emission}
\shortauthors{Faerber et al.}

\begin{document}

\title{Expansion Signatures in 35 \hii\ Regions traced by SOFIA \cii\ Emission}

\author[0009-0003-2745-5849]{T. Faerber}
\affiliation{Center for Gravitational Waves and Cosmology, West Virginia University, Chestnut Ridge Research Building, Morgantown, WV 26505, USA}
\author[0000-0001-8800-1793]{L.D. Anderson}
\affiliation{Center for Gravitational Waves and Cosmology, West Virginia University, Chestnut Ridge Research Building, Morgantown, WV 26505, USA}
\author[0000-0001-8061-216X]{M. Luisi}
\affiliation{Department of Physics, Westminster College, New Wilmington, PA 16172, USA}
\affiliation{Center for Gravitational Waves and Cosmology, West Virginia University, Chestnut Ridge Research Building, Morgantown, WV 26505, USA}
\author[0000-0002-0915-4853]{L. Bonne}
\affiliation{Stratospheric Observatory for Infrared Astronomy, Universities Space Research Association, NASA Ames Research Center, MS 232-11, Moffett Field, 94035
CA, USA}
\author[0000-0003-3485-6678]{N. Schneider}
\affiliation{I. Physikalisches Institut, Universit\"at zu K\"oln, Z\"ulpicher Str. 77, 50937 K\"oln, Germany}
\author[0000-0002-8351-3877]{V. Ossenkopf-Okada}
\affiliation{I. Physikalisches Institut, Universit\"at zu K\"oln, Z\"ulpicher Str. 77, 50937 K\"oln, Germany}
\author[0000-0003-0306-0028]{A. G. G. M. Tielens}
\affiliation{Leiden Observatory, Leiden University, Niels Bohrweg 2, 2333 CA Leiden, The Netherlands}
\affiliation{Department of Astronomy, University of Maryland, College Park, MD 20742, USA}
\author[0000-0003-2555-4408]{R. Simon}
\affiliation{I. Physikalisches Institut, Universit\"at zu K\"oln, Z\"ulpicher Str. 77, 50937 K\"oln, Germany}
\author[0000-0001-6205-2242]{M. R\"ollig}
\affiliation{Physikalischer Verein – Gesellschaft f\"ur Bildung und Wissenschaft,
Frankfurt am Main, Germany}
\affiliation{Institut f\"ur Angewandte Physik, Goethe-Universit\"at Frankfurt,
Frankfurt am Main, Germany}

%\author[0000-0002-3466-6164]{James M. Jackson}
%\affiliation{Green Bank Observatory}

\begin{abstract}

We analyze the expansion signatures of 35 \hii\ regions mapped in \cii\ 158\,\micron\ emission by the Stratospheric Observatory for Infrared Astronomy (SOFIA). The \cii\ emission primarily traces photodissociation regions (PDRs) at the transition between ionized and neutral gas. The brightness and narrow linewidth of \cii\ allow us to measure PDR expansion. Bubble-shaped regions often exhibit expansion, while irregular-shaped ones are less likely to. Of the 35 \hii\ regions, 12 ($\sim\!34\%$) exhibit clear expansion in position-velocity (PV) diagrams, making them expansion candidates (ECs), with an average expansion velocity of $\sim\!12.2$\,\kms. The remaining 23 regions show no clear expansion signatures, though they may still be expanding below detection limits. Blueshifted expansion is more common (eight ECs solely blueshifted; one redshifted; three both), with mean velocities of $\sim\!10.9$\,\kms\ (blueshifted) and $\sim\!13.2$\,\kms\ (redshifted). A comparison of our observations to spherical expansion models supports expansion in eight of 12 ECs. Estimated dynamical ages are 10 to 100 times shorter than the ionizing star lifetimes, in agreement with the results of previous studies. Of the 35 regions, 14 ($\sim\!40\%$) appear as \cii\ bubbles; nine of the 12 ECs are bubble-shaped. Thermal pressure likely drives expansion in M43, while stellar winds dominate in M17, M42, RCW 120, and RCW 79. For other ECs, available data do not allow a definitive conclusion. Larger samples and more information about ionizing sources are needed to refine our understanding of \hii\ region feedback and evolution.

\end{abstract}


\section{Introduction}
\label{sec:introduction}

\hii\ regions are ionized areas of the interstellar medium (ISM) surrounding high-mass OB spectral-type stars. They occur when dense clouds of cold, neutral molecular/atomic gas become ionized by strong ultraviolet radiation from young OB stars that reside within them \citep{Stromgren1939}. The prevalence, brightness, and angular size of \hii\ regions makes them natural laboratories for detailed studies of the impact of high-mass stars, such as triggered star formation and cloud disruption.

Because of their simple geometry, studies of \hii\ region expansion and triggered star formation have largely focused on so-called ``bubble'' \hii\ regions.  Approximately half of all identified \hii\ regions can be classified as bubbles based on their mid-infrared (MIR; 5-40 \micron) morphology, appearing annular or ring-shaped on a 2D map \citep{churchwell2006bubbling, anderson2012dust, deharveng2010observations}. We assume that the ring-like shape seen in a 2D projection of 3D real-space is also homogeneous along the third spatial axis along the line-of-sight in at least one direction, resulting in a semi-spherical or spherical shell. Such a shell has been observed in the Cygnus region, referred to as the ``Diamond Ring,'' which is assumed to represent a final stage of an expanding \cii\ bubble (Dannhauer et al., in prep).

This ring-like structure is generally observed in the transition region between the ionized gas of the \hii\ region and the predominantly molecular gas beyond it called the photodissociation region (PDR). Although PDRs lack significant ionized hydrogen, they do have ionized gas from atoms with ionization potentials $\leq 13.6$\,eV. The most common ion in PDRs is therefore ionized carbon (\cii), as C has an ionization potential of 11.3\,eV and is the fourth most abundant element. 

As the distance from the ionizing source increases, photons with energies $\geq11.3$\,eV are absorbed, resulting in a relatively thin ionized carbon shell. However, it is not the distance itself, but rather the column density of absorbing material that primarily determines the extent of absorption \citep{tielens1985photodissociation, hollenbach1999photodissociation, wolfire2022photodissociation}.

The \cii\ ($^2P_{3/2}\!\rightarrow\!^2P_{1/2}$) emission line at 158\,\micron\ is a commonly used spectral line for probing the spatial and kinematic properties of \hii\ regions \citep{stacey1991optical, anderson2019origin, pabst2020expanding, luisi2021stellar, bonne2022sofia, bonne2023sofia}. This line is one of the brightest emission lines in far-infrared spectra in the ISM, and is therefore a powerful tool for investigating PDR dynamics where ionized carbon is present among the neutral gas \citep{Tielens2005}. The choice of \cii\ emission as a tracer for analyzing \hii\ regions is driven by several factors. In addition to being a bright tracer of \hii\ region PDRs, \cii\ emission has a relatively long wavelength of $\sim\!$158 \micron\ \citep{Tielens2005}, allowing it to penetrate dense clouds of gas and dust that absorb radiation at shorter wavelengths, revealing regions of the ISM that may be otherwise hidden from our view. However, the \cii\ line is optically thick in bright, dense PDRs \citep{guevara2020c, kabanovic2022self} and therefore does not always trace the full extent of the PDR layer.

The use of heterodyne techniques at 158 \micron\ (THz frequencies) enables spectral resolving powers of up to $R\approx10^6$ (where $R=\frac{\nu}{\Delta\nu}=\frac{\Delta\lambda}{\lambda}$ is the spectral resolving power), corresponding to sub-\kms\ velocity resolutions, which is crucial for detecting fine-scale velocity structures indicative of expansion within PDRs. In contrast, optical and mid-infrared studies typically rely on slit-spectrum grating spectrometers with resolving powers between $R = 10^3 - 10^5$. While high-resolution systems such as TEXES \citep{lacy2002texes, zhu2005mass} have enabled similar analyses in the mid-infrared, such instrumentation is limited and rarely reaches the precision achievable in the radio regime. In the scope of this research, the \cii\ line allows us to search for kinematic signatures of expansion in the PDRs of a sample of \hii\ regions.

The physical expansion of \hii\ regions as they interact with their surrounding medium can influence the structure of the nearby interstellar environment. This expansion, driven by radiation and stellar winds from central OB stars, can compress surrounding gas and shape molecular cloud morphology. In some cases, this process may lead to secondary star formation along the edges of the expanding region, a phenomenon often referred to as ``triggered star formation.'' Proposed mechanisms include the ``collect-and-collapse'' scenario \citep{elmegreen1977sequential, weaver1977interstellar, lancaster2024geometry} and radiatively driven implosion (RDI) \citep{lefloch1997triggered, bertoldi1989photoevaporation}. Observational studies have debated the statistical significance of triggered star formation \citep{pomares2009triggered, deharveng2010observations, schneider2012cluster, dale2015dangers}, although individual cases, such as G24.47+0.49 \citep{saha2024direct}, W5 \citep{karr2003triggered}, and RCW 120 \citep{zavagno2006triggered}, show evidence of multi-epoch or localized triggered star formation. RDI and collect-and-collapse can also operate simultaneously, as suggested by simulations of RCW 120 by \cite{walch2015comparing}. However, the presence of young stars does not always imply triggering, as seen in the Rosette Nebula, where stellar clusters are not associated with local OB stars \citep{cambresy2013young}.

The morphology of an \hii\ region can influence whether or not signs of expansion are detectable in its PDR. If the region is homogeneous and has a bubble morphology, it is more likely for uniform expansion to be detected than if the region is irregular. Irregular expansion is likely due to inhomogeneities in the density of the medium surrounding the central ionizing star.

Observational evidence for the expansion of \hii\ regions has accumulated over the past two decades, particularly through velocity differences between the ionized gas and surrounding PDRs. The dominant drivers of this expansion—thermal pressure from photoionized gas and mechanical input from stellar winds—remain under active investigation. In some cases, expansion appears consistent with classical models of pressure-driven expansion, while in others, stellar wind feedback may be required to explain the observed kinematics. Observations with SOFIA have proven especially effective in tracing PDR dynamics, revealing expansion signatures in a number of well-studied regions \citep{pabst2020expanding, luisi2021stellar, tiwari2021sofia, beuther2022feedback, bonne2022sofia, bonne2023sofia}.

Most prior studies, however, have focused on individual regions and apply a range of different analysis methods, making it difficult to assess broader trends or compare expansion mechanisms across different environments. In this work, we aim to address this gap by applying a uniform methodology to a large sample of \hii\ regions observed in \cii. A detailed comparison between our results and previous studies is presented in Section~\ref{subsec:comparision}. The work done in this paper builds on the approach of \cite{luisi2021stellar}, expanding it to a larger sample of 35 regions to establish statistics on the expansion of \hii\ regions and explore the roles of thermal pressure and stellar wind pressure in driving any observed expansion.

%\subsection{\hii\ region Expansion}
\section{\hii\ Region Expansion}
\label{sec:hii_region_expansion}

\subsection{Thermal Expansion}
\label{subsec:thermal_exp}

The Str\"omgren radius, $R_{s,0}$, of an \hii\ region is defined as the radius from an ionizing UV source at which ionization and recombination are balanced \citep{Stromgren1939}. The initial stationary Str\"omgren radius of a ``dust-free'' \hii\ region is given by Equation~\ref{eq:R_{s,0}}:
\abovedisplayskip=0pt
\belowdisplayskip=0pt
\begin{equation}
    R_{s,0} = \left(\frac{3F^{*}}{4\pi n_{0}^{2} \alpha_{B}}\right)^{1/3},
    \label{eq:R_{s,0}}
\end{equation} 
\noindent where $n_0$ is the ambient density of the cold, neutral medium, $F^*$ is the total number of ionizing UV photons per second coming from the hydrogen ionizing source (OB star), and $\alpha_B = 2.6 \times 10^{-13}$ cm$^3$ s$^{-1}$ is the hydrogen recombination coefficient (to all levels above the ground level, assuming an electron temperature of $T_e\approx10^{4}$ K). If we take the range of $F^*$ to be from $1\times10^{46}-7.5\times10^{49}$ photons s$^{-1}$ (lower bound is typical for a B1 star, upper bound is typical for an O3 star; \cite{martins2005new}) and the range of ambient densities to be $n_0=100-10^4$ cm$^{-3}$ (star-forming regions do not have densities as low as 100 cm$^{-3}$, however, as we are concerned with expansion out of star-forming regions into surrounding giant molecular clouds (GMCs) we can use the density of GMCs, which can have densities as low as 100 cm$^{-3}$; \cite{Draine2011}), we find the theoretical range in $R_{s,0}$ to be from 0.01 pc to 6.17 pc.

%For a dusty \hii\ region, the initial Str\"omgren radius can be estimated by simply adding an exponential term  so that $R_{s, dust} \simeq R_{s,0} e^{-\tau/3}$, where $\tau$ is the UV optical depth along the line of sight due to absorption from dust \citep{franco1990formation}. 

\begin{figure*}[ht]
    \centering
    \includegraphics[width=\textwidth]{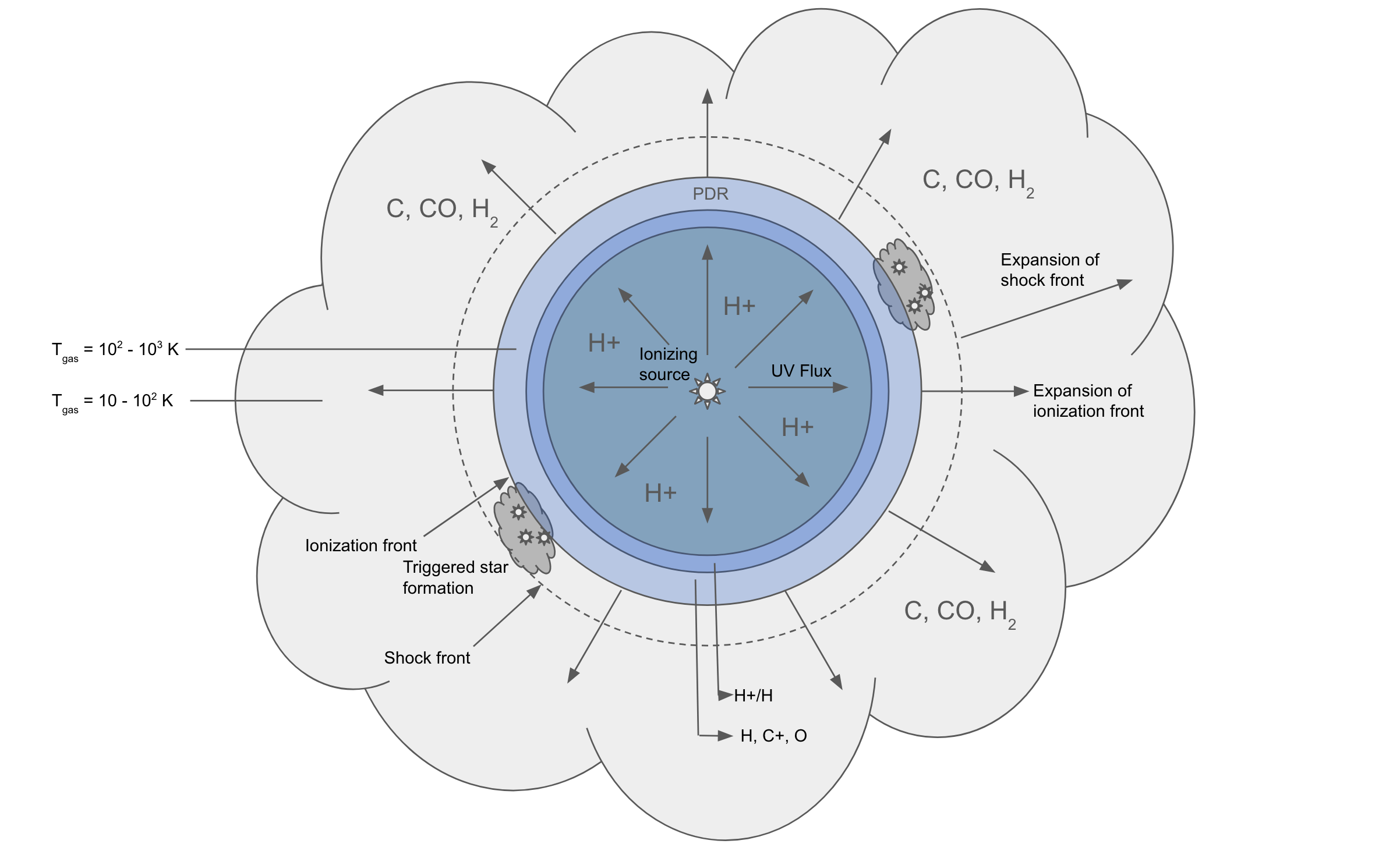}
    \caption{Schematic of the expanding ionization front of an \hii\ region with triggered star formation at the edge of the expanding shell. This figure is not to scale, as the ionization front is very thin, about $\sim\!10^{-3}$\,pc, compared to the size of the \hii\ region at $\sim\!1$\,pc \citep{raga2012universal}. In reality, the structure of an \hii\ region is much more complex with substructures in the region and a reverse shock front. For the purposes of this study we stick with this simplified model of an \hii\ region's structure to interpret our observations.}
    \label{fig:hii_region_diagram}
\end{figure*}

Figure~\ref{fig:hii_region_diagram} presents a schematic diagram of an expanding \hii\ region. The central ionizing source is depicted as a star at the center. Surrounding it is a region of ionized hydrogen (\hii), shown as a dark blue circle, where ionized hydrogen is more prevalent than other ionized species. This \hii\ region has an overpressure that drives a shock front into the surrounding molecular cloud, sweeping up gas into a dense shell. The inside of this shell is illuminated by far-UV (6–13.6 eV) radiation, creating a PDR boundary, shown as the outer light blue ring. As the distance from the ionizing source increases, UV photons with energies $E \geq 13.6$ eV are absorbed, but species like carbon, which has an ionization potential below 13.6 eV, can still be ionized. The small dark clouds along the outer edge of the PDR front represent sites where triggered star formation might occur as the PDR expands into the surrounding molecular cloud. A neutral shell is trapped between the ionization front and the shock front.

The time that it takes for an \hii\ region to reach its initial Str\"omgren radius is given by the recombination timescale, $t_r = (\alpha_{B}*n_0)^{-1}$, where $\alpha_B$ is the ``Case B'' hydrogen recombination coefficient as in Equation~\ref{eq:R_{s,0}} \citep{spitzer1968diffuse}.

As an \hii\ region evolves, this Str\"omgren sphere expands beyond its initial Str\"omgren radius. This expansion is driven by increased thermal pressure of the ionized gas or the pressure of the hot plasma created by the stellar wind from the ionizing source(s). Initially, this expansion is faster than the sound speed in the ambient medium, causing a shock front that propagates either in front of, with, or eventually falls behind the ionization front \citep{franco2000evolution}. 

An \hii\ region's radius at time $t$ due to thermal expansion as a function of its initial Str\"omgren radius is given by the relation from \cite{spitzer1968diffuse}:
\begin{equation}
    R_{\rm therm} = R_{s,0} \left( 1+\frac{7}{4} \frac{c_it}{R_{s,0}}\right)^{4/7}\,
    \label{eq:R_thermal}
\end{equation}
\noindent where $c_i$ is the sound speed in the warm, ionized medium (typically around 10\,\kms). $R_{\rm therm}$ represents the inner boundary of the PDR shell. If we differentiate Equation~\ref{eq:R_thermal} with respect to $t$, we get the thermal expansion rate for an \hii\ region:
\begin{equation}
    \frac{dR_{\rm therm}}{dt} = v_{\rm therm} =c_i \left( 1+ \frac{7}{4} \frac{c_i t}{R_{s,0}}\right)^{-3/7}.
    \label{eq:v_thermal}
\end{equation}

%If we allow $\dot{r}$ to evolve as $t\rightarrow\infty$, we notice that $r\rightarrow\infty$ while $\dot{r}\rightarrow0$. 

In theory, as the HII region expands, its internal pressure drops and as it nears the pressure of the surrounding medium, causing the shock wave to become subsonic and the expansion to stall \citep{franco1989expansion}. %To account for this, \cite{raga2012analytic} takes the solution for the expansion rate of an \hii\ region from \cite{spitzer1968diffuse} a step further, introducing a term that accounts for post-shock material  (gas that has been passed over by the expanding shock front of the material, leaving it with higher density, pressure, and temperature and the pre-shock gas still unperturbed by the shock front)}. The solution for thermal expansion velocity now becomes:

%\begin{equation}
 %   v'_{thermal} = \left[ c_i \left( \frac{R_{s,0}}{R_{\rm therm}} \right)^{3/4} - \sigma %\left( \frac{R_{s,0}}{R_{\rm therm}} \right)^{3/4} \right],
  %  \label{eq:v_thermal_post_shock}
%\end{equation}
%\noindent where $\sigma=c_o^2/c_i^2$ and $c_o$ is the sound speed in the cold, neutral medium. When $\sigma=0$, we have the relation from Equation~\ref{eq:v_thermal}. The sound speed in each medium is given by the relation $c = \sqrt{\frac{\gamma kT_e}{m_p}}$,  where $T_e$ is the temperature of the \hii\ region and $m_p$ is the mass of a proton.} In the warm, ionized medium we can take $T_e\approx10^4$ K which gives a sound speed of  $c_i=10.74$\,\kms. In the cold, molecular medium we can take $T_e\approx10$ K, which gives a sound speed of $c_o=0.34$\,\kms. This gives a value for sigma of $\sigma\approx10^{-3}$.}

If and when pressure equilibrium is eventually reached between the \hii\ region and the surrounding molecular cloud, the expansion of the ionization front can stall at a stagnation radius given by the following relation from \cite{bisbas2015starbench}:
\begin{equation}
    R_{\rm stag} = R_{s,0}\left(\frac{2c_i}{c_o}\right)^{4/3}\,,
    \label{eq:R_stag}
\end{equation}
\noindent where $c_o$ is the sound speed in the cold, neutral medium (typically around 0.3 \kms.) Using sound speeds of $c_i \approx 10.5$\,\kms\ and $c_o \approx 0.3$\,\kms\, the theoretical range for $R_{\rm stag}$ is between 3.7\,pc and 792.5\,pc.

%\noindent where $c_o$ is the sound speed in the cold, neutral medium (typically around 0.3 \kms.) Using sound speeds of $c_i = 10.74$\,\kms\ and $c_o = 0.34$\,\kms\, the theoretical range for $R_{\rm stag}$ is between 3.68\,pc and 792.49\,pc. 

%After the ionization boundary of an \hii\ region reaches this stagnation radius, simulations from \cite{williams2018classical} show that the boundary goes through strongly damped oscillation about $R_{\rm stag}$.  

If we plug $R_{\rm stag}$ from Equation~\ref{eq:R_stag} into the expansion rate solution from \cite{spitzer1968diffuse} for $r(t)$ and solve for $t$, we get the time that it will take the expansion of an \hii\ region to stagnate, $t_{\rm stag}$:
\begin{equation}
    t_{\rm stag} = \frac{4}{7} \frac{R_{s,0}}{c_i} \left[\left(\frac{R_{\rm stag}}{R_{s,0}}\right)^{4/7} -1\right].
    \label{eq:t_stag}
\end{equation}
\noindent Using the ranges in $R_{s,0}$ and $R_{\rm stag}$ defined above, $t_{\rm stag}$ is seen to have a theoretical range from around 15,000 yr to 3 \myr. The OB stars powering \hii\ regions have typical lifetimes of approximately 5-10 \myr, which is slightly longer on average than the theoretical range in stagnation times.

{\subsection{Stellar wind driven expansion}
\label{subsec:stellar_wind_exp}

Expansion in \hii\ regions can also be driven by stellar winds from the ionizing source(s), generating cavities around massive stars called ``wind-blown bubbles'' (WBBs). \cite{weaver1977interstellar} considers the interaction of strong stellar winds with the interstellar medium in the early and intermediate stages of stellar evolution. The radius of the ``cold shell,'' $R_2$, as defined in \cite{weaver1977interstellar}, is the radius of the outer shock front where the transition occurs between the hot and cold medium. This can be considered analogous to the outer edge of the PDR as defined by Equation~\ref{eq:R_{s,0}}, therefore, we will refer to this radius as $R_{\rm wind}$. The evolution of $R_{\rm wind}$ over time is given as:
\abovedisplayskip=0pt
\belowdisplayskip=0pt
\begin{equation}
    R_{\rm wind} = \left(\frac{250}{308\pi}\right)^{1/5}L_w^{1/5}\rho_0^{-1/5}t^{3/5},
    \label{eq:R_wind}
\end{equation}where $L_w$ is the mechanical luminosity of the stellar wind, $\rho_0$ is the density of the ambient medium, and $t$ is time. Given a reasonable range of input parameters (molecular hydrogen with $n=100$–$10^4$ cm$^{-3}$; 
$t$ = $10^4$ yr - 10 Myr; $L_w = 10^{34} - 10^{37}$ erg s$^{-1}$), the theoretical range for $R_{\rm wind}$ ranges from 0.1 pc to 244.0 pc. \footnote{We use the range of $L_w$ values found in the literature from \cite{ellerbroek2013RCW36} and \cite{rosen2014gone}, which represent minima and maxima for the sample of \hii\ regions in this study.} The expansion velocity of this interface over time is given by integrating Equation~\ref{eq:R_wind} with respect to $t$, giving:
\abovedisplayskip=0pt
\belowdisplayskip=0pt
\begin{equation}
    \frac{dR_{\rm wind}}{dt} = v_{\rm wind} = \left(\frac{250}{308\pi}\right)^{1/5}\frac{3}{5}L_w^{1/5}\rho_0^{-1/5}t^{-2/5}.
    \label{eq:v_wind}
\end{equation}
\noindent Using the same range for the input parameters shows that stellar wind can drive expansion with velocities of up to tens of\,\kms, with a theoretical upper limit of $\sim\!60$\,\kms\ (though expansion velocities this high have not been observed).

\section{Data}
\label{sec:data}

The Stratospheric Observatory for Infrared Astronomy (SOFIA, decommissioned on September 29, 2022) was an airborne facility, representing a collaboration between NASA and the German Aerospace Center (DLR) \citep{Krabbe2008}. The observatory was equipped with a suite of instruments including the German Receiver for Astronomy at Terahertz Frequencies (GREAT) which was used to conduct high-resolution spectroscopic observations \citep{heyminck2012great}. This was eventually replaced by upGREAT, an updated version of GREAT, during SOFIA Cycle 4 in 2016 \citep{risacher2018upgreat}. upGREAT introduced mid-sized heterodyne arrays to GREAT, increasing its mapping speed by a factor of $\sim\!$10 \citep{Guesten2014}.  The angular resolution for \cii\ observations from SOFIA is 14.1\arcsec, while the velocity resolution is 0.04\,\kms\ (raw) and 0.2\,\kms\ (resampled for FEEDBACK sources, see below). 

In this paper, we search for signs of expansion in all 35 \hii\ regions observed by SOFIA in \cii\ emission. We list the regions used in this study in Table~\ref{tab:region_information} along with literature references (when previous literature for the source exists) and several notes about the sources and observations. 

The 35 \hii\ regions have a range of properties.  Their ionizing sources range from single O stars (O9 – O6), to small clusters ($\sim\!10$ O stars) to starburst regions. Four of the regions are located in the Large Magellanic Cloud (LMC) or the Small Magellanic Cloud (SMC). These regions are N160, N44, N66, and N79. 
Of the 35 regions observed, 13 of them (Cygnus X, M16, M17, NGC 6334, NGC 6334IV, NGC 6334V, NGC 7538, RCW 120, RCW 36, RCW 49, RCW 79, W40 and W43) were observed as part of the FEEDBACK project \citep{schneider2020feedback}. The FEEDBACK project is a SOFIA Legacy Program aimed at understanding how massive stars interact with their environment by mapping \cii\ and \oi\ emission in Galactic \hii\ regions to study feedback-driven gas dynamics. Details of the FEEDBACK observations are presented in \cite{schneider2020feedback}.

All but seven of the 35 \hii\ regions have distance estimates in the literature. For the seven that do not (G081+036, G083+936, G287+814, G301+138, G316+796, G317+426, and G320+088), we estimate distances using a Monte Carlo kinematic distance method from \cite{wenger2018kinematic}, using the Galactic rotation curve from \cite{reid2014trigonometric}. This method takes the known Galactic longitude, latitude, and observed velocity of each source and compares the observed velocity to the expected velocity at different distances along the line of sight, as predicted by the Galactic rotation model. Since the parameters of the Galactic rotation curve and the observed velocity have uncertainties, the Monte Carlo method generates many random variations (resamples) of these parameters within their uncertainty ranges, producing a large set of possible distances. Each set of resampled values yields a new distance estimate, and after many iterations, a probability distribution of distances is built. For sources within the inner Galaxy, where two possible distances (near and far) can result from the same velocity, the most probable distance is determined by analyzing the distribution of all calculated distances and selecting the peak value, with uncertainties defined by the range of distances around this peak. 

\begin{table*}
    \centering
    \caption{Information for 35 \hii\ regions observed by SOFIA in \cii\ emission}
    \label{tab:region_information}
    \resizebox{\textwidth}{!}{
        \begin{tabular}{lccccccccc}
        \hline
            Region & RA & Dec & SOFIA map size [n $\times$ m] & \hii\ region radius [\arcsec] & Distance [kpc] & log($F^*$) & Bubble? & Literature \\
        \hline
        Cygnus & 20h38m30s & 42d13m00s & $2505 \times 2695$ & 480.7 & $\sim\!1.5$\tablenote{\footnotesize \cite{rygl2012parallaxes}} & 49.58 & No & \cite{schneider2023ionized}; \cite{bonne2023unveiling} \\
        G082+036 & 20h32m21s & 43d41m14.8s & $930 \times 660$ & 376.8 & $3.49 (+0.96 / -0.91)$\tablenote{\footnotesize \cite{reid2014trigonometric}; \cite{wenger2018kinematic}} & \nodata & No & \nodata \\
        G083+936 & 20h45m38.9s & 44d14m52.3s & $630 \times 439$ & 220.0 & 0.88 $\pm$ 0.02\ $^b$ & \nodata & Yes & \nodata \\
        G287+814 & 10h45m53.7s & $-$59d57m05.1s & $750 \times 300$ & 80.6 & $1.52 (+0.37 / -0.96)$; $3.65 (+0.94 / -0.53)$ $^b$ & \nodata & No & \nodata \\
        G301+138 & 12h35m36s & $-$63d02m33.2s & $1200 \times 600$ & 54.8 & $3.29 (+0.51 / -0.67)$; 5.64 (+0.39 / -0.75) $^b$ & \nodata & Yes & \nodata \\
        G316+796 & 14h45m19.7s & $-$59d49m34.5s & $675 \times 540$ & 311.8 & 2.43 $\pm$ 0.42; 9.66 $\pm$ 0.45 $^b$ & \nodata & Yes & \nodata \\
        G317+426 & 14h51m36.5s & $-$60d00m25.1s & $585 \times 300$ & 71.7 & $14.53 (+0.59 / -0.63)$ $^b$ & \nodata & No & \nodata \\
        G320+088 & 15h14m34.5s & $-$58d10m35.9s & $450 \times 240$ & 54.9 & $2.64 (+0.48 / -0.19); 10.30 (+0.20 / -0.57)$ $^b$ & \nodata & No & \nodata \\
        M8 & 18h03m46.6s & $-$24d22m19.5s & $255 \times 300$ & 429.8 & 1.25 $\pm$ 0.1\tablenote{\footnotesize \cite{damiani2019wide}; \cite{tothill2008lagoon}} & 48.63 & No & \cite{tiwari2019observational} \\
        M16 & 18h18m30s & $-$13d45m00s & $2295 \times 2130$ & 1124.4 & 2.10\tablenote{\footnotesize \cite{mcbreen1982high}} & 49.66 & No & \cite{garcia2006faint}; \cite{karim2023sofia} \\
        M17 & 18h20m43.8s & $-$16d09m41.5s & $1050 \times 570$ & 1200.0 & 1.98\tablenote{\footnotesize \cite{xu2011trigonometric}} & 49.77 & Yes & \cite{povich2007multiwavelength}; \cite{lim2020surveying}; \cite{guevara2020c} \\
        M20 & 18h02m29.3s & $-$23d01m27.9s & $525 \times 480$ & 550.4 & 1.67\tablenote{\footnotesize \cite{lynds1985optical}} & 48.80 & No & \cite{garcia2006faint} \\
        M42 & 05h34m40.5s & $-$05d37m51.9s & $4050 \times 2820$ & 1749.6 & 0.44\tablenote{\footnotesize \cite{jeffries2007distance}} & 48.63 & Yes & \cite{pabst2020expanding} \\
        M43 & 05h35m30.0s & $-$05d17m00.0s & $675 \times 570$ & 200.0 & 0.44\tablenote{\footnotesize \cite{jeffries2007distance}} & 47.00 & Yes & \cite{pabst2020expanding}; \cite{guevara2020c} \\
        Mon R2 & 06h07m46.8s & $-$06d22m57.1s & $900 \times 468$ & 241.8 & $0.78 (+0.40 / -0.37)$\tablenote{\footnotesize \cite{trevino2019dynamics}} & 48.96 & No & \cite{trevino2019dynamics} \\
        N19 & 18h18m29s & $-$13d40m00s & $509 \times 645$ & 500.0 & 2.10\tablenote{\footnotesize \cite{mcbreen1982high}} & 48.29 & Yes & \cite{xu2019effects} \\
        N160 & 05h39m35s & $-$69d39m00s & $975 \times 240$ & 100.0 & $49.59 \pm 0.54$\tablenote{\footnotesize \cite{pietrzynski2019distance}} & 50.23 & No & \cite{martin2008mid} \\
        N44 & 05h22m06s & $-$67d58m00s & $675 \times 360$ & 100.0 & $49.59 \pm 0.54$\tablenote{\footnotesize \cite{pietrzynski2019distance}} & 49.00 & No & \cite{barman2022study} \\
        N66 & 00h59m00s & $-$72d10m00s & $1050 \times 285$ & 100.0 & 62.44 $\pm$ 0.47\tablenote{\footnotesize \cite{graczyk2020distance}} & 50.32 & No & \cite{geist2022ionization} \\
        N79 & 04h51m54s & $-$69d23m30s & $540 \times 180$ & 100.0 & $49.59 \pm 0.54$\tablenote{\footnotesize \cite{pietrzynski2019distance}} & 49.00 & No & \cite{ochsendorf2017star} \\
        NESSIE-A & 16h41m08s & $-$47d07m10.9s & $720 \times 480$ & 109.0 & 3.10\tablenote{\footnotesize \cite{goodman2014bones}} & 49.00 & Yes & \cite{ragan2014giant}; \cite{jackson2024absorption} \\
        NGC 1977 & 05h35m30.0s & $-$04d46m42.9s & $2025 \times 2520$ & 540.0 & 0.44\tablenote{\footnotesize \cite{jeffries2007distance}} & 45.48 & Yes & \cite{pabst2020expanding} \\
        NGC 2074 & 05h39m09s & $-$69d30m28s & $390 \times 132$ & 100.0 & $49.59 \pm 0.54$\tablenote{\footnotesize \cite{pietrzynski2019distance}} & 50.25 & No & \cite{fleener2009massive} \\
        NGC 6334 & 17h20m31.2s & $-$35d51m37.8s & $240 \times 180$ & 119.0 & 1.75\tablenote{\footnotesize \cite{russeil2016ngc}} & 48.76 & No & \cite{jackson1999photodissociation}; \cite{carral2002detection}; \cite{russeil2016ngc} \\
        NGC 6334IV & 17h20m17.7s & $-$35d54m29.2s & $210 \times 240$ & 131.5 & 1.75\tablenote{\footnotesize \cite{russeil2016ngc}} & 45.48 & No & \cite{jackson1999photodissociation}; \cite{carral2002detection}; \cite{russeil2016ngc} \\
        NGC 6334V & 17h19m57.3s & $-$35d57m09.8s & $240 \times 180$ & 119.0 & 1.75\tablenote{\footnotesize \cite{russeil2016ngc}} & 45.48 & No & \cite{jackson1999photodissociation}; \cite{carral2002detection}; \cite{russeil2016ngc} \\
        NGC 7538 & 23h13m47.8s & 61d30m01.9s & $2250 \times 1040$ & 573.9 & 2.65\tablenote{\footnotesize \cite{moscadelli2014multiple}} & 49.64 & Yes & \cite{beuther2022feedback}; \cite{werner1979infrared} \\
        RCW 120 & 17h12m22.2s & $-$38d26m54.3s & $1275 \times 1002$ & 500.1 & 1.70\tablenote{\footnotesize \cite{kuhn2019kinematics}} & 48.29 & Yes & \cite{anderson2010physical}; \cite{zavagno2010star}; \cite{luisi2021stellar} \\
        RCW 36 & 08h59m28.8s & $-$43d45m28.7s & $1470 \times 560$ & 810.0 & 1.09\tablenote{\footnotesize \cite{damiani2019wide}} & 48.44 & No & \cite{bonne2022sofia} \\
        RCW 49 & 10h23m50.7s & $-$57d45m50s & $2400 \times 2100$ & 1333.0 & 2.30\tablenote{\footnotesize \cite{belloni1994rosat}}; $\sim\!6 – 7$\tablenote{\footnotesize \cite{benaglia2013high}} & 49.48 & No & \cite{whitney2004glimpse}; \cite{tiwari2021sofia} \\
        RCW 79 & 13h40m27.7s & $-$61d40m27.6s & $1200 \times 600$ & 477.5 & 3.9\tablenote{\footnotesize \cite{bonne2023sofia}} & 49.78 & Yes & \cite{zavagno2006triggered} \\
        S235 & 05h41m01.2s & 35d51m16s & $750 \times 1440$ & 438.3 & 1.60\tablenote{\footnotesize \cite{anderson2019origin}} & 47.56 & Yes & \cite{anderson2019origin} \\
        W40 & 18h31m28.7s & $-$02d08m22.3s & $1380 \times 1020$ & 1206.9 & 0.50\tablenote{\footnotesize \cite{comeron2022extended}} & 48.18 & Yes & \cite{10.1093/pasj/psy115} \\
        W43 & 18h47m30s & $-$02d00m00s & $129 \times 2170$ & 346.6 & 5.5\tablenote{\footnotesize \cite{bania19973he}; \cite{balser2001vla}} & 51.00 & No & \cite{luong2011w43} \\
        W51 & 19h23m50s & 14d30m40.5s & $252 \times 144$ & 16.6 & 1.70\tablenote{\footnotesize \cite{koo1997hi}} & 51.00 & No & \cite{lim2019surveying} \\
        \hline
        \end{tabular}
    }
\end{table*}

%\end{sidewaystable*}

%\tablenote{\footnotesize
%\cite{rygl2012parallaxes}; \cite{schneider2023ionized}; \cite{reid2014trigonometric}; \cite{wenger2018kinematic}; \cite{bonne2023unveiling}; \cite{damiani2019wide}; \cite{tothill2008lagoon}; \cite{tiwari2019observational}; \cite{mcbreen1982high}; \cite{garcia2006faint}; \cite{karim2023sofia}; \cite{xu2011trigonometric}; \cite{povich2007multiwavelength}; \cite{lim2020surveying}; \cite{lynds1985optical}; \cite{pabst2020expanding}; \cite{trevino2019dynamics}; \cite{xu2019effects}; \cite{pietrzynski2019distance}; \cite{martin2008mid}; \cite{barman2022study}; \cite{graczyk2020distance}; \cite{geist2022ionization}; \cite{ochsendorf2017star}; \cite{goodman2014bones}; \cite{ragan2014giant}; \cite{jackson2024absorption}; \cite{fleener2009massive}; \cite{russeil2016NGC }; \cite{carral2002detection}; \cite{moscadelli2014multiple}; \cite{beuther2022feedback}; \cite{werner1979infrared}; \cite{kuhn2019kinematics}; \cite{anderson2010physical}; \cite{zavagno2010star}; \cite{luisi2021stellar}; \cite{bonne2022sofia}; \cite{belloni1994rosat}; \cite{benaglia2013high}; \cite{whitney2004glimpse}; \cite{tiwari2021sofia}; \cite{bonne2023sofia}; \cite{comeron2022extended}; \cite{10.1093/pasj/psy115}; \cite{bania19973he}; \cite{balser2001vla}; \cite{luong2011w43}; \cite{koo1997hi}; \cite{lim2019surveying}.
%}

\section{Methods}
\label{sec:methods}

This study uses both position-velocity (PV) diagrams and velocity residual maps to search for expansion signatures in the \cii\ data. A PV diagram is a two-dimensional plot showing the intensity of spectral line emission as a function of spatial position and velocity along a chosen path. These paths are defined with start and end points on the moment-0 map of the \hii\ region, and are therefore perpendicular to the line of sight. %To create a PV diagram, one extracts the emission along a defined path in the two spatial axes of a datacube (which has two spatial axes and one spectral axis). If a strip with a finite width is used, the emission can be summed within this region.

One signature of expansion in a PV diagram is a semi-ellipse with the end-points of the semi-minor axis representing the systemic velocity. For uniform expansion, the PV diagram will resemble a full ellipse. For those \hii\ regions where an expansion signature is observed, we define a semi-ellipse by eye that approximates the signature seen in the PV diagram \citep[as in][]{luisi2021stellar}.

For each of the 35 regions in the study, we create 16 PV diagrams using paths at evenly-spaced angles through the \hii\ region centers.  For 21 of the 35 \hii\ regions, the center of the region is taken from the WISE catalog of Galactic \hii\ regions \citep{anderson2014wise}. For the remaining 14 regions (N79, N160, NESSIE-A, NESSIE-A sub-bubble, NGC 6334, NGC 1977, N44, N19, NGC 2074, G083+936, RCW 36, M17, N66 and M43) we define a center that better matches the \cii\ emission.
%by drawing a circle that enclosed the \cii\ moment-0 map emission and then use the center of that circle. 
 We define the path widths for each \hii\ region to be large enough that the resultant PV diagrams have high signal to noise but are not so large as to mix signals from different areas of the maps.
 %These 16 individual PV diagrams are used to study expansion along as many axes in each \hii\ region in the sample, allowing us to determine an average global expansion velocity and identify axes along which expansion may not be uniform.   
 Figure~\ref{fig:m17_output} shows two of the 16 PV diagrams taken along different paths across the moment-0 map of M17. The red areas overlaid on the moment-0 maps in the right panels show the region from which the PV diagram in the corresponding left panel is extracted. The PV x-axis, starting at zero, runs from the left to the right of the red area along its longest spatial axis (the shorter spatial axis is the one that is summed).

\begin{figure*}[ht]
    \centering
    \includegraphics[width=\textwidth]{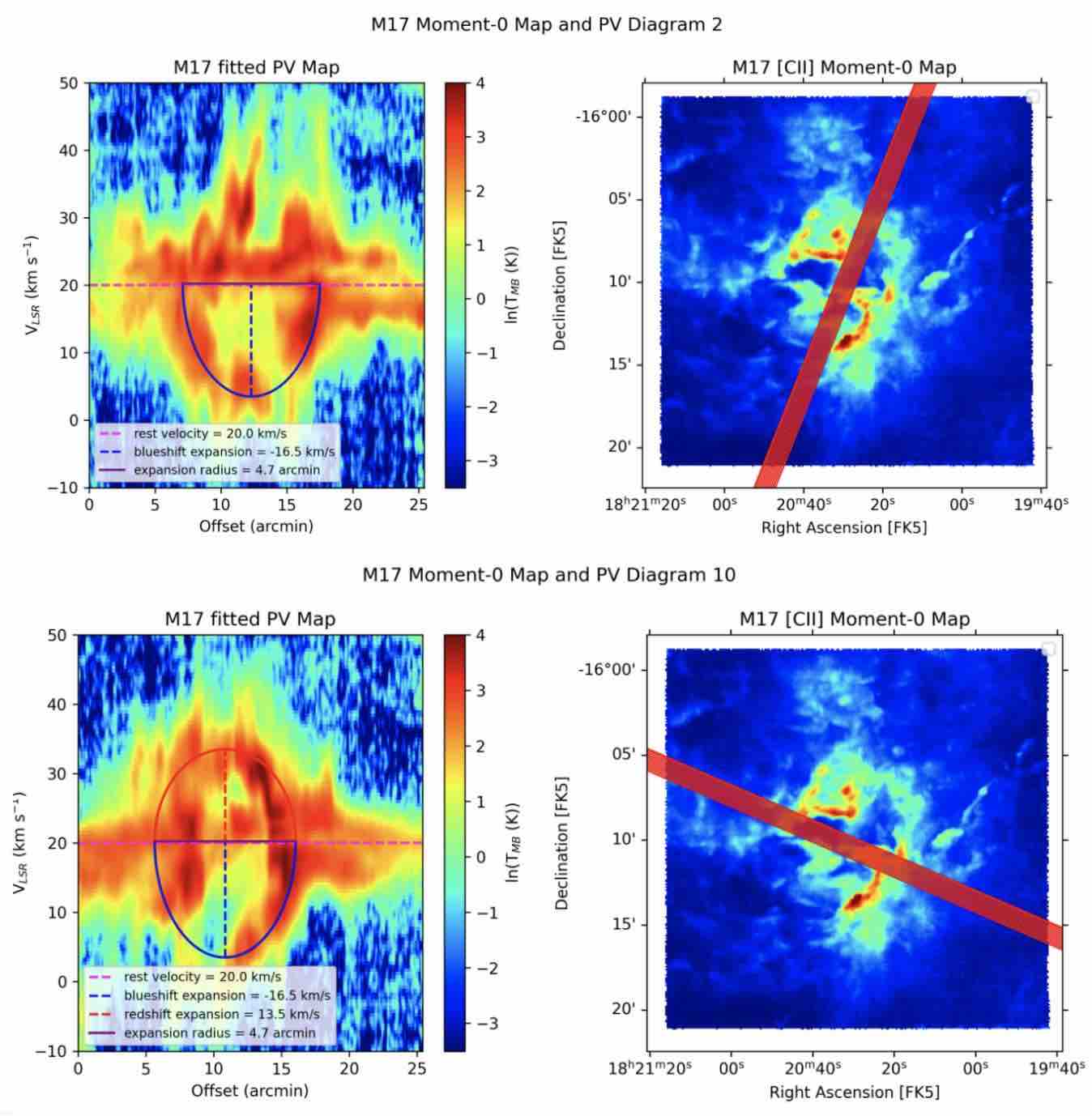}
    \caption{Two PV diagrams for M17. The left column shows the PV diagrams, and the right column shows moment-0 maps of the areas that the corresponding PV diagrams are extracted from. The x-axis in the first PV diagram represents distance from left to right along the red area in the moment-0 map. PV diagram 2 shows a blueshifted shell expanding at $\sim\!-16.5$ \kms\ with no redshifted counterpart. PV diagram 10 displays redshifted and blueshifted shells, expanding at $\sim\!13.5$ \kms\ and $\sim\!-16.5$ \kms, respectively.}
    \label{fig:m17_output}
\end{figure*}

%For regions where we detect a signal consistent with expansion in at least one of the 16 PV diagrams with a detection threshhold of $\leq\!1.5\sigma$...

For regions with visual identification of expansion in at least one of the 16 PV diagrams, we produce velocity residual maps. For this analysis, we first generate an observed velocity map of the region, where we bin the datacube in RA and Dec into $n\times m$ subcubes (where the length and width of each subcube is the same). These subcubes are much larger than the beamsize of SOFIA when observing at 158 \micron, which is around 14\arcsec. We choose the size of the subcubes based on the similar analysis performed by \cite{luisi2021stellar}, where subcube size is selected to maximize the signal-to-noise ratio of the \cii\ expansion signal. In each of these subcubes we create an average \cii\ spectrum of the observed emission and remove a fitted (using non-linear least squares optimization) Gaussian at the systemic velocity by subtracting it from the observed spectrum. The result is a rest-velocity subtracted spectrum where any residual peak in intensity could represent other velocity components in the subcube, such as expansion signatures. Figure~\ref{fig:spec_subtraction} shows this process carried out on a spectrum from one of the subcubes in RCW 120 containing an expansion signal. For blueshifted expansion we search for residual velocity components less than the subtracted systemic velocity, while for redshifted expansion we search for residual velocity components greater than the subtracted systemic velocity. We then generate a suite of predicted velocity maps for a uniformly expanding sphere with a range of expansion velocities and expansion centers.  On a 2D map, redshifted and/or blueshifted expansion of a sphere, when observed separately, will have the largest absolute value (manifesting as the largest offset from the systemic velocity) toward the center of the sphere, decreasing towards the edges until reaching the systemic velocity. 

\begin{figure*}[ht]
    \centering
    \includegraphics[width=0.55\textwidth]{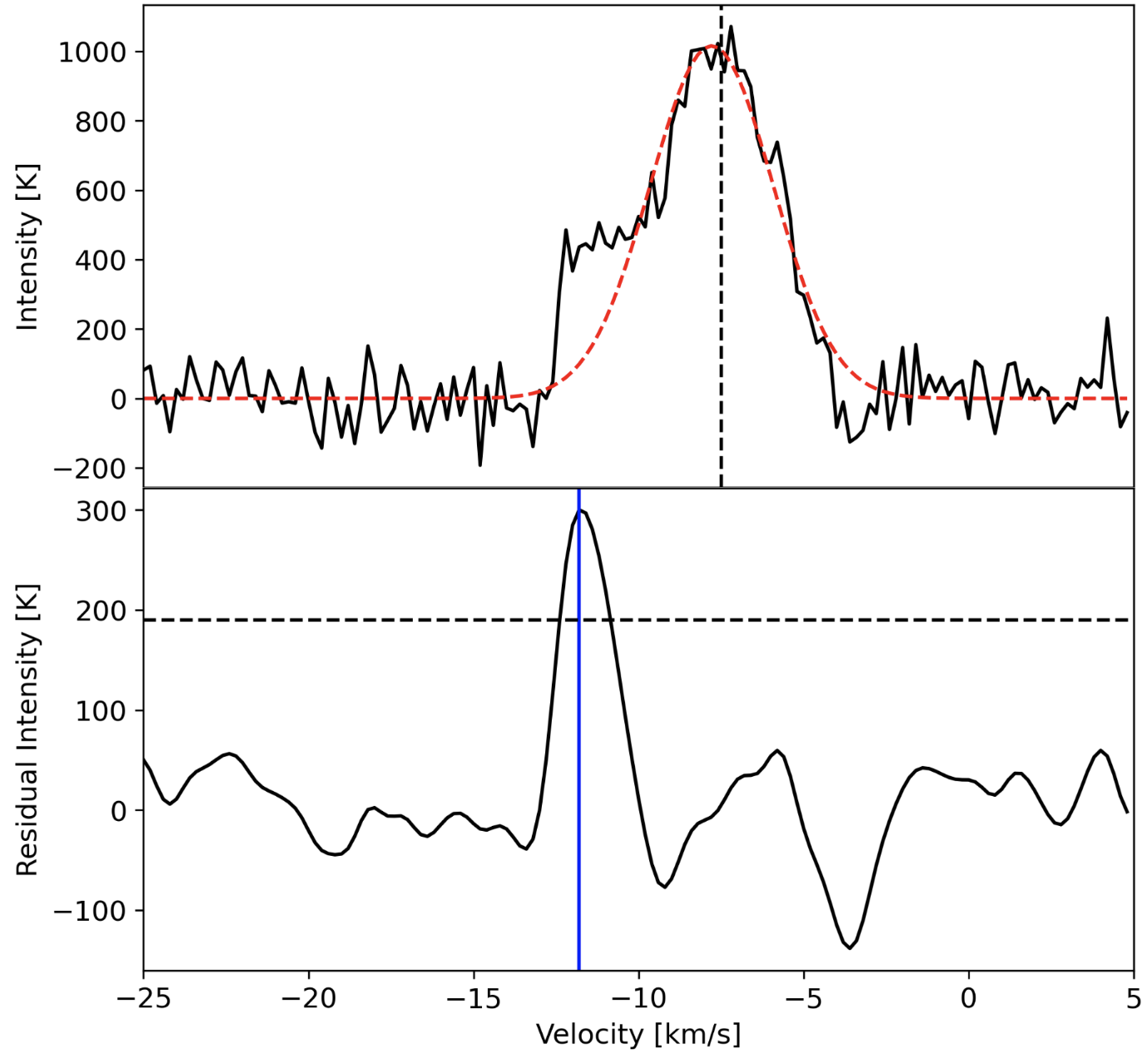}
\caption{Example of the spectral subtraction process used to identify expansion signatures in RCW 120. The top panel shows the observed \cii\ spectrum (black) extracted from a subcube, along with the fitted Gaussian (red dashed Gaussian) around the systemic velocity (vertical black dashed line). The bottom panel presents the residual spectrum after subtracting the fitted Gaussian, highlighting any remaining velocity components. A significant residual feature with a flux density greater than three times the rms noise at a blueshifted velocity relative to the systemic velocity (shown by the vertical solid blue line) is visible, indicating a possible expansion signature.}
    \label{fig:spec_subtraction}
\end{figure*}

Finally, we subtract the predicted from the observed velocity maps for all generated models and identify the model that provides the lowest overall mean of the absolute value of all residuals produced by the model. We display our method for RCW 120 in Figure~\ref{fig:RCW 120_vel_map}, which shows that the model of blueshifted expansion that best-fits the observations is one centered on the \cii\ emission (shown as contours) with an expansion velocity of $\sim\!16.5$\,\kms. \cite{luisi2021stellar} used the same method on an incomplete field of RCW 120 and found an expansion value of 14\,\kms.

\begin{figure*}[ht]
    \centering
    \includegraphics[width=0.75\textwidth] {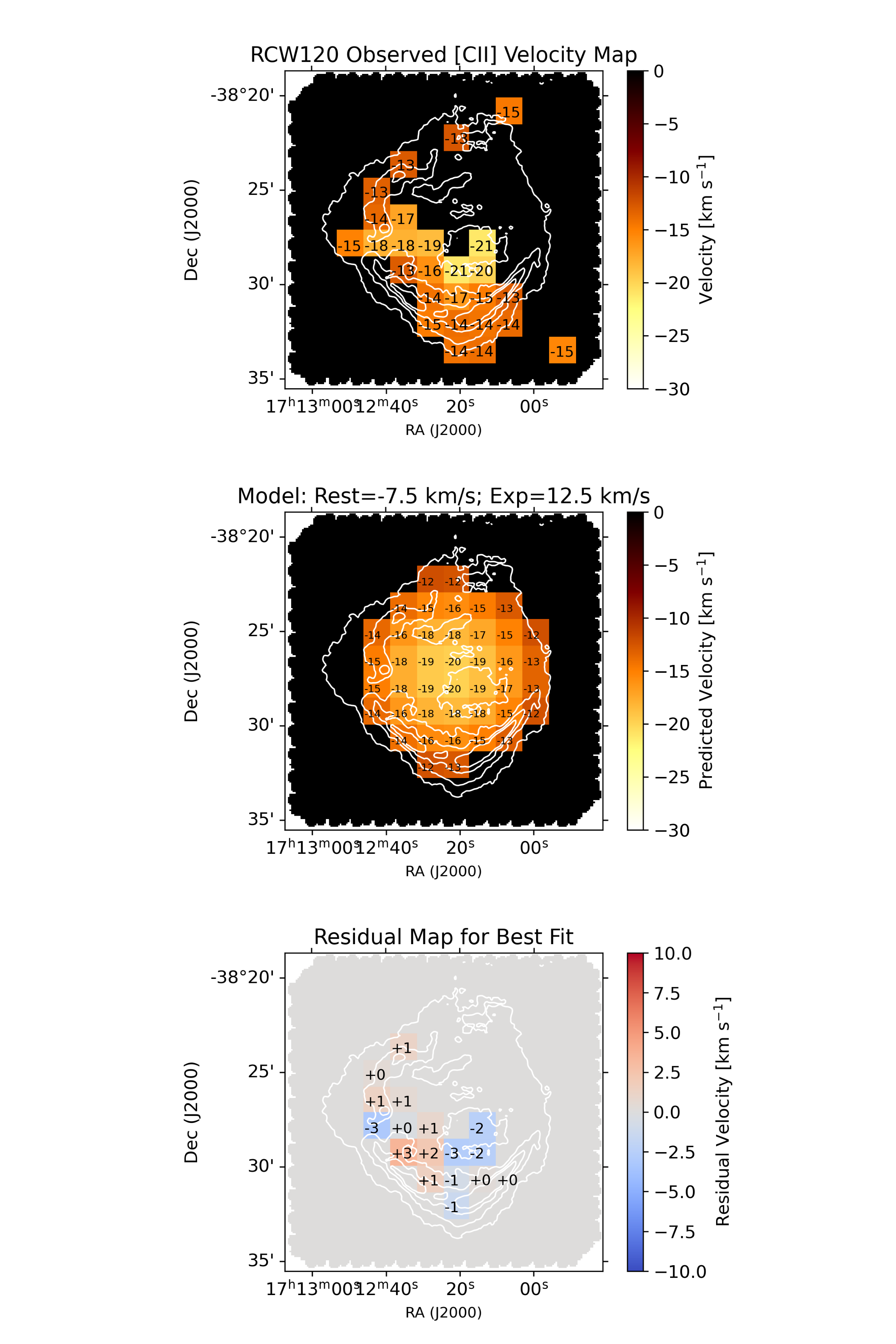}
    \caption{Velocity residual map analysis for RCW 120. \textit{Top:} Observed \cii\ bluesihfted velocity map for RCW 120, with expansion velocities in each subcube determined through the process outlined in Figure~\ref{fig:spec_subtraction}. \textit{Middle:} Predicted \cii\ velocity map for specified systemic velocity, expansion velocity, center of expansion and radius of expansion (systemic velocity and radius of expansion are defined as fixed parameters - expansion velocity and center of expansion are free parameters where a range of values is explored). \textit{Bottom:} Velocity residual map. The three subcubes with the largest magnitude of their residual are removed, as they change based on the region of the spectrum (minus the fitted rest Gaussian) that we search for a signal and cannot be fully trusted. For example, if we reduce the velocity cutoff from -10 \kms\ to -9 \kms\, the values in these subcubes for the observed map change from -11 \kms\ to -10 \kms.}
    \label{fig:RCW 120_vel_map}
\end{figure*}

\section{Results}
\label{sec:results}

%\subsection{Sample Statistics}
%\label{subsec:sample_statistics}

\subsection{Position-velocity Diagrams}
\label{subsec:pv_diagrams}
Of the 35 \hii\ regions studied in this paper, 12 exhibit expansion in at least one of their PV diagrams ($\sim\!34\%$ of the sample). Figure~\ref{fig:m17_output} shows two axes for M17 that show expansion signals, with PV diagram 2 showing only a blueshifted signal, while PV diagram 10 shows both a redshifted and a blueshifted signal. Table~\ref{tab:expansion_parameters} shows the global expansion parameters for these 12 expansion candidates (ECs), along with expansion velocity values from the literature (when available). Of the 12 ECs, only G316+796 displays solely redshifted expansion with no blueshifted signature while eight display only blueshifted expansion with no redshifted signature, and three show both redshifted and blueshifted expansion. The average expansion rate for all 12 ECs is $\sim\!12.2$\,\kms. The average blueshifted expansion velocity of the ECs is $\sim\!10.9$\,\kms\ while the average redshifted expansion velocity is $\sim\!13.2$\,\kms. The average percent of PV axes showing an expansion signature for all 12 ECs is 63\%. As is expected, all four of the ECs that display expansion in 100\% of their PV diagrams appear as bubbles in \cii\ emission (Table~\ref{tab:expansion_parameters}). Of the 12 ECs, nine of them are bubbles (75\%). On average, 72\% of the PV cuts display expansion in the nine ECs that have bubble morphologies. On the other hand, only 34\% of the PV cuts in the non-bubble ECs display expansion. Of the entire sample of 35 \hii\ regions, 14 are bubbles (40\%; see Table~\ref{tab:region_information}).

%\begin{sidewaystable*}
\begin{table*}
\centering
\tiny
\caption{Expansion Candidate Global Parameters}
\label{tab:expansion_parameters}
\begin{tabular}{lcccccccc}
\hline
Region & \% \footnote{The second column of Table~\ref{tab:expansion_parameters} (titled \%) shows the percent of PV cuts (out of 16 total for each region) that display an expansion signature. If the column says 100, then all 16 of the PV cuts display an expansion signature.} & Bubble? & Expansion Radius \footnote{For the columns titled ``Expansion Radius'', ``Blueshift Expansion'' and ``Redshift Expansion'', the values are given as such: (minimum value, maximum value) mean value $\pm$ standard deviation.} & Expansion (lit) & Blueshift Expansion & Redshift Expansion & Blue Age & Red Age \\
 &  &  & [pc] & [\kms] & [\kms] & [\kms] & [\myr] & [\myr] \\
\hline
G083+936 & 100 & Y & $(0.06, 0.15)~0.09 \pm 0.03$ & \nodata & ($-7.8, -9.3$) $-8.9 \pm 0.5$ & $(11.9, 13.4)~12.6 \pm 0.6$ & $0.01 \pm \lesssim0.01$ & $0.01 \pm \lesssim0.01$ \\
G316+796 & 50 & Y & $(1.84, 2.47)~2.05 \pm 0.23$ & \nodata & \nodata & $(12.2, 13.4)~13.0 \pm 0.4$ & \nodata & $0.16 \pm 0.02$ \\
M17 & 81 & Y & $(1.96, 2.88)~2.50 \pm 0.28$ & \nodata & $(-16.0, -19.5)~-17.5 \pm 1.3$ & $(13.0, 13.5)~13.2 \pm 0.2$ & $0.14 \pm 0.02$ & 0.21 $\pm$ 0.01 \\
M42 & 100 & Y & $(1.41, 2.14) 1.68 \pm 0.24$ & $-13$\footnote{Pabst et al. 2019} & $(-13.4, -14.0)~-13.8 \pm 0.2$ & \nodata & $0.12 \pm 0.02$ & \nodata \\
M43 & 100 & Y & $(0.06, 0.23)~0.13 \pm 0.06$ & $-6$\footnote{Pabst et al. 2019} & $(-3.5, -4.0)~-3.6 \pm 1.5$ & \nodata & $0.04 \pm 0.01$ & \nodata \\
N19 & 63 & N & (2.75, 4.58) 3.91 $\pm$ 0.61 & $-4$ (in prep) & $(-8.2, -8.8)~-8.3 \pm 0.2$ & \nodata & $0.47 \pm 0.08$ & \nodata \\
NESSIE-Aa\footnote{There are no associated errors on the parameters for NESSIE-Aa because there is only one PV diagram showing expansion. Therefore there is no standard deviation, which is how the errors were determined.} & 13 & Y & $(0.18, 0.18)~0.18$
 & \nodata & $(-16.4, -16.4)~-16.4$ & \nodata & 0.01 & \nodata\\
NGC 7538 & 44 & Y & $(1.31, 2.00)~1.51 \pm 0.25$ & $\sim\!$-10 \footnote{Beuther et al. 2022} & $(-11.1, -15.9)~-14.6 \pm 1.6$ & \nodata & $0.11 \pm 0.03$ & \nodata \\
RCW 120 & 100 & Y & $(1.73, 2.97)~2.17 \pm 0.40$ & $\sim\!$-15 \footnote{Luisi et al. 2021} & $(-14.0, -16.0)~-15.3 \pm 0.6$ & \nodata & $0.14 \pm 0.03$ & \nodata\\
RCW 36 & 19 & N & $(0.98, 1.21)~1.10 \pm 0.09$ & $\sim\!-5.2$ \footnote{Bonne et al. 2021} & $(-7.8, -9.0)~-8.5 \pm 0.5$ & \nodata & $0.13 \pm 0.01$ & \nodata \\
RCW 79 & 63 & Y & $(7.94, 11.91)~9.64 \pm 1.16$ & $\leq$ 25 \footnote{Bonne et al. 2023} & $(-8.5, -14.8)~-11.6 \pm 2.6$ & $(13.5, 14.7)~14.1 \pm 0.5$ & $0.72 \pm 0.07$ & $0.85 \pm 0.26$ \\
W40 & 19 & N & $(0.43, 0.55)~0.50 \pm 0.05$ & \nodata & $(-11.4, -12.3)~-12.0 \pm 0.4$ & \nodata & $0.04 \pm 0.01$ & \nodata \\
\hline
\end{tabular}  
\end{table*}
%\end{sidewaystable*}

\subsection{Velocity Residual Maps}
\label{subsec:velocity_residual_maps}

Table~\ref{tab:velocity_residual_maps} shows the results of the velocity residual map analysis for eight of the 12 ECs where the emission has a bubble morphology, appearing annular in its \cii\ moment-0 map, and either a radius on the sky of at least $2.5\arcmin$ or a peak \cii\ moment-0 intensity of at least 400 K\,\kms\ (G083+936, G316+796, N19 and RCW 36 are too faint and/or compact). 
%These sample restrictions are used as they are found to produce the most reliable models for the sources observed with regards to expansion center and expansion velocity (for example, if the model produced has a large physical offset from the observed expansion center in the PV diagram it is considered unreliable). 
The second column labeled ``$n\times m$'' gives the binning parameters used for the spatial axes (see Section~\ref{sec:methods}). The $v_{\rm r, map}$ and $v_{\rm b, map}$ columns give the values for the expansion velocity of the best-fit model for redshifted and blueshifted shells, respectively. The associated errors for these values can be inferred from the standard deviation plot shown in the right panel of Figure~\ref{fig:snr_robustness}. Using the determined velocity of expansion and the observed S/N, the standard deviation can be determined. These standard deviations can also be used to determine the size of the error bars for the x-axis in Figure~\ref{fig:pv_vs_res}. The $\langle v \rangle _{\rm r, PV}$ and $\langle v \rangle _{\rm b, PV}$ columns indicate the average expansion values (across 16 PV axes per region) from the position-velocity diagram analysis outlined in Section~\ref{subsec:pv_diagrams} for comparison also for redshifted and blueshifted shells, respectively. PV axes that did not show signs of expansion are excluded from the average, which can lead to artificially inflated expansion values in regions where only a few axes exhibited expansion.

\begin{table*}
    \centering
%    \tiny
    \caption{Velocity Residual Map Results}
    \label{tab:velocity_residual_maps}
%    \resizebox{\textwidth}{!}{%
        \begin{tabular}{lcccccccc}
        \hline
            Region & $n\times m$ & Subcube size & $v_{\rm r, map}$ & $\langle v \rangle _{\rm r, PV}$ & log(Red S/N) & $v_{\rm b, map}$ & $\langle v \rangle _{\rm b, PV}$ & log(Blue S/N) \\
             & & [\arcsec] & [\kms] & [\kms] & & [\kms] & [\kms] & \\
            \hline
            M17 & $16\times16$ & 34.99 & $10.0\pm0.8$ & $13.2\pm0.2$ & 2.04 & $14.0\pm0.4$ & $17.5\pm1.3$ & 2.15 \\
            M42 & $16\times23$ & 133.13 & \nodata & \nodata & \nodata & $12.0\pm0.6$ & $13.8\pm0.2$ & 1.56\\
            M43 & $18\times19$ & 149.72 & \nodata & \nodata & \nodata & $7.0\pm2.5$ & $3.6\pm1.5$ & 1.17\\
            NESSIE-Aa & $18\times22$ & 26.66 & \nodata & \nodata & \nodata & $10.0\pm1.1$ & 16.4 \footnote{No associated error because there is only one PV diagram showing expansion.} & 0.96\\
            NGC 7538 & $16\times16$ & 24.37 & \nodata & \nodata & \nodata & $11.0\pm1.3$ & $14.6\pm1.6$ & 0.43\\
            RCW 120 & $16\times16$ & 83.53 & \nodata & \nodata & \nodata & $12.5\pm1.0$ & $15.3\pm0.6$ & 0.51\\
            RCW 79 & $16\times16$ & 22.50 & $14.5\pm0.5$ & $14.1\pm0.5$ & 1.14 & $18.0\pm0.2$ & $11.6\pm2.6$ & 1.08\\ 
            W40 & $22\times22$ & 128.18 & \nodata & \nodata & \nodata & $13.0\pm0.5$ & $12.0\pm0.4$ & 1.46\\   
            \hline
        \end{tabular}
%    }
\end{table*}

Only two of the ten expansion signatures with associated errors in Table~\ref{tab:velocity_residual_maps} fall within the error bars of their corresponding PV analyses: the redshifted expansion in RCW 79 and the blueshifted expansion for M43; suggesting one-to-one agreement of the two methods for these signals. The remaining nine signatures fall outside their PV-derived uncertainties, but still produced well-constrained expansion models based on the \cii\ moment-0 morphology. The average expansion velocity from the ten velocity residual map analyses is 12.8\,\kms. Two maps model redshifted shells (M17 and RCW 79) with values of 10.0\,\kms\ and 14.5\,\kms, respectively. The remaining eight maps model blueshifted shells (M17, M42, M43, NESSIE-A, NGC 7538, RCW 120, RCW 79, and W40), with an average modeled blueshifted expansion velocity of 12.2\,\kms. To test the robustness of this analysis, we repeat it on simulated data, as described in Section~\ref{subsec:vel_res_robust_test}. Figure~\ref{fig:pv_vs_res} shows the expansion velocity derived from our PV analysis on the y axis versus the expansion velocity estimated from the velocity map on the x axis. As indicated by the one-to-one line in Figure~\ref{fig:pv_vs_res}, the estimated expansion velocities for almost all regions fall within $5\sigma$ for the two approaches considered.

\begin{figure*}[ht]
    \centering
    \includegraphics[width=0.8\textwidth]{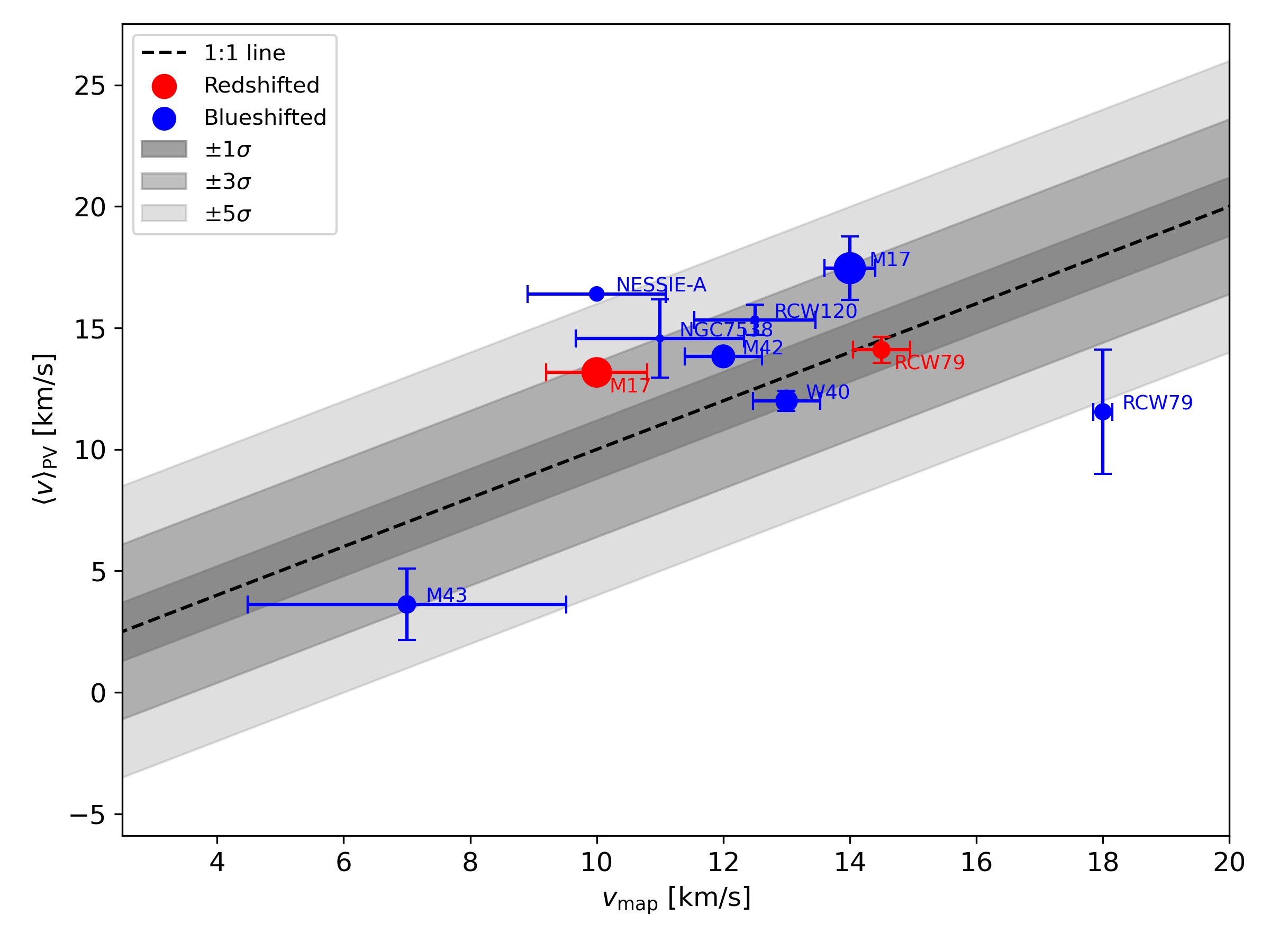}
    \caption{Comparison of estimated expansion velocities. The y-axis shows the expansion velocity determined through our PV analysis versus the expansion velocity determined through our velocity map analysis on the x-axis. The x-axis errors are determined from the simulations producing the standard deviation plot in the right panel of Figure~\ref{fig:snr_robustness}. The black dashed line represents a one-to-one ratio of the two expansion velocities, so that regions falling closer to this line represent those with better agreement between the PV and velocity map analyses. The gray shaded areas show where the observations are within $1\sigma$, $3\sigma$ and $5\sigma$ of the one-to-one line, where $\sigma=1.2$ \kms\ is estimated as the standard deviation of the residuals between the PV and residual map velocities. The size of the filled circles scales as the size of the S/N of the observed signal, such that larger markers indicate an observation with a larger S/N and smaller markers indicate an observation with a smaller S/N. All values used to generate the plot can be found in Table~\ref{tab:velocity_residual_maps}.}
    \label{fig:pv_vs_res}
\end{figure*}

\section{Discussion}
\label{sec:discussion}

\subsection{Comparison with Previous Observations}
\label{subsec:comparision}
Several previous studies have identified expansion signatures in \hii\ regions, primarily based on velocity offsets between the ionized gas and surrounding PDRs. Our analysis builds on this body of work by providing a uniform comparison across a large sample.

\citet{roshi20058} observed carbon recombination lines toward 18 ultracompact \hii\ regions and identified dense PDRs in 11 sources. In nine cases, they found a consistent velocity offset of $\sim$3.3\,\kms\ between the PDR and ionized gas, which they interpreted as evidence of expansion. \citet{kirsanova2017gas} studied 14 regions in CS(2–1) and $^{13}$CO(1–0), detecting a 1.2\,\kms\ offset in G183.35$-$0.58, which they interpreted as expansion of the \hii\ region into the molecular cloud. In follow-up work, \citet{kirsanova2020pdr} observed S235 in \cii, interpreting the velocity structure as further evidence of expansion.

Several studies using \cii\ observations with SOFIA have reported expansion in individual \hii\ regions. \citet{pabst2020expanding} reported expansion velocities of $\sim$13\,\kms\ in M42, $\sim$6\,\kms\ in M43, and $\sim$1\,\kms\ in NGC 1977, concluding that stellar winds likely dominate the expansion in M42, while thermal pressure may drive expansion in M43 and NGC 1977. These observations have been extended in subsequent work by \citet{pabst2021ii, pabst2022158, pabst2024multiline}. \citet{luisi2021stellar} observed an expansion velocity of $\sim$15\,\kms\ in RCW 120 and attributed the expansion to stellar winds. Prior work by \citet{zavagno2006triggered} suggested that the expansion of RCW 120 may be triggering star formation along the southwestern edge of the region. \citet{tiwari2021sofia} identified expansion in RCW 49 with a similar velocity ($\sim$13\,\kms), supported by evidence of stellar wind-driven expansion from diffuse X-ray emission.

More recent SOFIA observations have reported expansion velocities of $\sim$5\,\kms\ in RCW 36 \citep{bonne2022sofia} and NGC 7538 \citep{beuther2022feedback}. \citet{bonne2023unveiling} observed simultaneous blueshifted and redshifted \cii\ emission in RCW 79, identifying an expansion velocity of up to $\sim$25\,\kms. This represents the first detection of both blueshifted and redshifted expansion signatures in the same region using \cii. \citet{figueira2017star} suggested that expansion in RCW 79 may be responsible for triggering star formation at the interface between the \hii\ region and the surrounding molecular material.

Finally, \citet{saha2024direct} used ALMA and the VLA to observe G24.47+0.49 in HCO$^+$(1–0) and radio continuum emission. They identified a 9\,\kms\ expanding ring of HCO$^+$ surrounding the \hii\ region and multiple collapsing dense cores along the expansion front, providing direct observational evidence of triggered star formation.

While these studies provide important context for understanding feedback-driven dynamics, they generally focus on individual regions. In contrast, our sample includes 35 \hii\ regions analyzed using a consistent method, allowing us to examine expansion signatures statistically across a wide range of morphologies and evolutionary stages. In several cases, our results corroborate previously measured expansion velocities (e.g., M42, M43, RCW 120, RCW 79), while in others, we provide new measurements for previously unstudied regions.

\subsection{Blueshift-dominated Expansion}
\label{sec:blue_exp}

Blueshifted expansion is more common than redshifted expansion in our sample of 12 ECs. Of the 15 total expansion signatures, 11 are blueshifted and only four are redshifted. We test the statistical significance of this asymmetry using a binomial test under the null hypothesis that blueshifted and redshifted signatures occur with equal probability (50\%). The resulting p-value of 0.081 indicates that this asymmetry is statistically significant at the 5\% level. Selection bias may influence the observed asymmetry, though the sample size of 12 ECs is too small to determine meaningful population statistics. The FEEDBACK program primarily targets well-known regions of massive star formation, which may be more likely to be ``blister'' \hii\ regions like Orion, with dense background molecular clouds.  Such a geometry would make them easier to detect at optical wavelengths, but the molecular material would inhibit redshifted expansion.
%However, the sample may still effectively probe expansion along the path of least resistance, which preferentially occurs in half a hemisphere.  It is also worth mentioning that the original sample size of 35 \hii\ regions is reduced to only 12 ECs through PV analysis, so the sample size is certainly too small to make any meaningful assumptions about the nature of all \hii\ regions. Any claims made in this section are merely about the sample of 12 ECs that we identify through this work.}

\subsection{Radial Expansion Asymmetry}
\label{subsec:expansion_asymmetry}

Eight of the 12 ECs ($\sim\!67\%$) do not display signs of expansion along every axis sampled. Estimated expansion velocities along the 16 axes differ, which we interpret as expansion asymmetry. We show this radial expansion asymmetry in Figure~\ref{fig:RCW79_vs_M42}, which defines the expansion velocities along 16 different axes in the region RCW 79 (left panel).   

\begin{figure*}[ht]
    \centering
    \includegraphics[width=\textwidth]{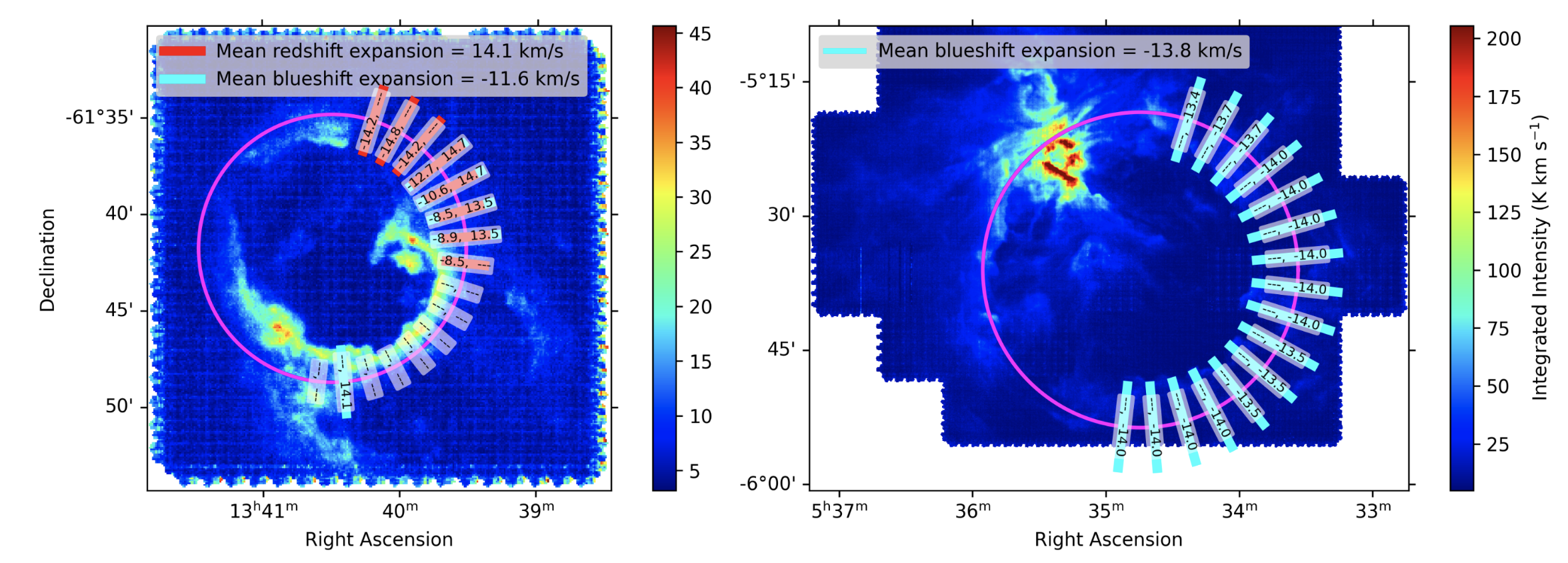}
    \caption{\cii\ Moment-0 map of RCW 79 (left) and M42 (right) with expansion velocities listed for each axis of expansion. For RCW 79, the first listed velocity is the measured redshift along the corresponding axis while the second listed number is the measured blueshift along the axis. The imaginary line connecting each pair of expansion velocity values to the center of the red circle corresponds to the axis (across the entire region) along which the aforementioned expansion velocities are observed. ``-\,-\,-'' represents no expansion seen along that axis of expansion. The red and cyan bars represent the relative magnitude of the observed redshifted and blueshifted expansion, respectively. RCW 79 and M42 are chosen to be compared as the blueshifted values for RCW 79 exhibit the most asymmetry of any EC in the sample, while for M42 they show the least.}
    \label{fig:RCW79_vs_M42}
\end{figure*}

We choose RCW 79 as an example because it has the largest expansion asymmetry of any EC in the sample. The blueshifted expansion values of RCW 79 range from $-8.5\,\kms$ to $-14.8\,\kms$. There are also six axes stretching from the northeast to the southwest of the region (where the material is likely denser and therefore the intensity of \cii\ emission is higher, as a higher intensity of emitting radiation from ionized gas corresponds to a larger column density of the emitting species) that do not exhibit any detectable expansion. Our observations are consistent with a non-uniform density in the surrounding medium where the expansion is fastest and easiest to detect along axes pointing towards less dense surroundings. The right panel of Figure~\ref{fig:RCW79_vs_M42} shows the same plot for M42, where the expansion is found to be the most uniform of any EC in the sample ($-13.4\,\kms$ to $-14.00\,\kms$ with 100\% of the PV diagrams showing blueshifted expansion signatures). It is possible that M42 displays a more uniform density in its surrounding medium, leading to a more uniform expansion than that seen in regions such as RCW 79.

\subsection{Non-Expanding \hii\ Regions}
\label{subsec:non_expanding_regions}

Of the 35 \hii\ regions observed in this study, 23 do not exhibit any detectable signs of expansion in their \cii\ PV diagrams. While this might initially suggest that these regions have reached pressure equilibrium, we tested this possibility by estimating their theoretical stagnation radii, $R_{\rm stag}$.

For each of the 23 non-expanding regions, we compiled values of the ionizing photon flux, $F^*$, from the literature and calculated $R_{\rm stag}$ following Equation~\ref{eq:R_stag}. In all cases where $F^*$ is available, we found that the radius of the \hii\ region is smaller than its predicted stagnation radius by at least a factor of 10. This indicates that these regions have not yet reached pressure equilibrium with the surrounding ISM and are therefore not expected to have stagnated.

The lack of detected expansion signatures could be the result of several factors. First, expansion velocities may be too low to distinguish them from the systemic velocity given our spectral resolution. Second, expansion may be occurring asymmetrically or along lines of sight not probed by our PV diagrams. Finally, the surrounding environment may be too dense or clumpy to support a clearly defined expanding shell. Despite the absence of observable expansion features, the fact that these regions have not yet reached their stagnation radii suggests that they are likely still undergoing some expansion, potentially in a slow or highly non-uniform fashion.

\subsection{Non-uniform Velocity Residual Maps}
\label{subsec:non_uniform_vel_maps}

As can be seen in the bottom panel of Figure~\ref{fig:RCW 120_vel_map}, we do not always attain a best-fit model with a velocity residual map where the residuals are all zero. Although RCW 120 shows expansion in all 16 of the axes probed with PV diagrams (see Table~\ref{tab:expansion_parameters}), the velocity residual map still shows some subcubes with larger residual values than others. The velocity residuals are larger in the northwest of the region, where \citet{luisi2021stellar} suggests that the bubble is bursting open into the surrounding ISM. This could point towards velocity components in the ionized gas along the line of sight that do not follow uniform expansion in the PDR. Such components could represent the erosion of molecular clouds as suggested by \cite{bonne2023sofia}. The column density of \cii\ is also smaller in the center of the region than in the limb-brightened edges, which results in a smaller S/N in the center. This will result in a less robust extraction of the expansion signal, which could be influencing our residual map. For this reason, when defining our best-fit model using the sum of all residuals, we remove three of the subcubes from the observed map in Figure~\ref{fig:RCW 120_vel_map} near the center of the region, which give values highly dependent on the velocity used as a cutoff threshold to search for a residual expansion signal after rest Gaussian is subtracted.

\subsection{Velocity Residual Map Analysis Robustness Test}
\label{subsec:vel_res_robust_test}

To test the robustness of the results from our velocity residual map analysis carried out in Section~\ref{subsec:velocity_residual_maps} we re-run the velocity residual map analysis on simulated data with a range of signal-to-noise ratios and expansion velocities. We allow the expansion velocity to vary from 0 to 18\,\kms\ by steps of 2\,\kms\ while we allow the S/N to range from $-2$ to 2 in log-space by steps of 0.5 using Equation~\ref{eq:snr_lenz} from \cite{lenz1992errors}:

\begin{equation}
    \rm{S/N} = 0.7 \times \frac{\it T_L}{{\rm rms}} \left(\frac{\Delta \it V}{2.35} \right)^{1/2},
    \label{eq:snr_lenz}
\end{equation}
\noindent where $\Delta V$ is the velocity resolution of the observations, around 0.5\,\kms, $T_L$ is the brightness of the expansion signal, and the rms is the spectral noise of the observation. The most possibly influential fixed parameter is the simulated line width for both the rest and expansion signal, which is set to 5\,\kms\ for both. This linewidth is selected as it is characteristic of the observed emission lines in 12 ECs.

We run the simulation 100 times for each combination of the two free parameters. We show the average difference between the simulated and fitted expansion velocities for the explored free parameter space in Figure~\ref{fig:snr_robustness}. In this figure, the left panel shows the absolute difference between the simulated and fitted expansion velocities while the right panel color bar shows the absolute difference as a percent of the simulated expansion velocity. We expect that when the S/N value of the expansion signal is higher and the simulated expansion velocity is higher (the upper right corner of both graphs in Figure~\ref{fig:snr_robustness}), our results will be more reliable and the pixels will have lower values in this area. This relationship is seen in both of the plots. The vertical lines shown in both plots indicate the S/N of the observations with reliable fits from this analysis. The cyan lines represent the blueshifted signals while the red lines represent the redshifted signals. The observed S/N values are estimated as the average S/N of the redshifted/blueshifted expansion signal in the brightest 50\% of subcubes within the radius defined in Section~\ref{sec:methods}.

Table~\ref{tab:velocity_residual_maps} summarizes the results of the velocity residual map fitting for each region with a successful fit and also shows the S/N for the observed expansion signals. Using the range in expansion velocities from this table along with the tabulated S/N values, we can look at Figure~\ref{fig:snr_robustness} and see that most of our observations fall in the upper right-hand corner of both plots with S/N values greater than $\textrm{log}_{10}(0.5)=3$ and a sample average expansion velocity of $\sim\!12.2$\,\kms.

\begin{figure*}[ht]
    \centering
    \includegraphics[width=\textwidth]{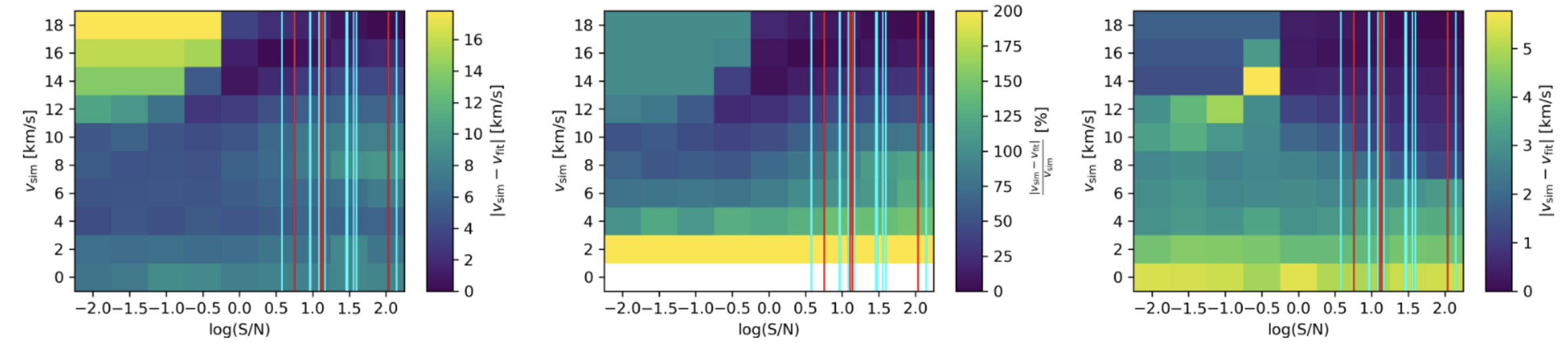}
    \caption{Parameter space for velocity residual map analysis simulations with simulated line width of 5\,\kms. The y-axis shows the probed expansion velocity parameter space while the x-axis shows the S/N parameter space in log space. The vertical lines shown in both plots indicate the S/N of the observations with reliable fits from this analysis. The cyan lines represent the blueshifted signals while the red lines represent the redshifted signals. \textit{Left:} The color at each pixel in the left panel corresponds to the average absolute velocity offset over 100 iterations between the simulated expansion velocity and fitted expansion velocity for a simulated model with the corresponding expansion velocity and S/N. \textit{Center:} The color at each pixel in the center panel corresponds to the offset between the simulated and fitted expansion velocities as a percentage of the simulated expansion velocity over 100 iterations. \textit{Right:} The color at each pixel in the right panel corresponds to the standard deviation of the offset between the simulated and fitted expansion velocities over 100 iterations.}
    \label{fig:snr_robustness}
\end{figure*}

For the 23 regions without a detected expansion signal, it is likely that these non-detections could be due to a combination of three different factors. The expansion velocity could be so low (or zero, producing no expansion signal) that its expansion signal blends with the systemic velocity signal. How much this interferes with detecting the expansion signal will be a function of the line width (which we have set as a fixed parameter in the simulations) and the expansion velocity, since a smaller line width and larger expansion velocity will have a clearer expansion signal, while a wider line width and smaller expansion velocity will result in a blended rest and expansion signal. It is also possible that the expansion signal is low enough and/or the noise could be high enough that the S/N is not large enough to produce a clear expansion signal. Finally, a region with an irregular shape and non-uniform expansion will be harder to fit with the velocity residual map analysis than a region appearing as a bubble with uniform spherical expansion. It could be a combination of all three of these factors that results in no detection of expansion signals in 23 of the 35 regions in the sample ($\sim\!66\%$).

\subsection{PDR Radii and Dynamical Lifetimes}
\label{subsec:pdr_radii_and_dynamical_lifetimes}

The observed PDR radii are shown in Table~\ref{tab:stagnation_times} along with stagnation radii and theoretical radii for both stellar wind and thermal pressure driven expansion. All of the observed PDR radii fall below the theoretical range for the stagnation times. It is therefore unlikely but not impossible that we observe effects of stagnation during the ionizing source’s lifetime.

If we assume that each EC begins its lifetime with a PDR radius of zero and make the simplifying assumption that the expansion velocity of the PDR front remains uniformly spherical (such that the physical radius along the line of sight is the same as the radius tangent to the observer's line of sight on the sky) and constant throughout the lifetime of the \hii\ region, we can work backwards to estimate an upper limit on the dynamical age, $t_{\rm dyn}$, for each EC. This is an upper limit because the expansion velocity that we observe now should theoretically be the slowest that the expansion has been during the lifetime of the region, starting its lifetime with its fastest expansion and eventually slowing to the current observed expansion velocity. Since expansion with respect to the systemic velocity of each \hii\ region is studied with redshift and blueshift separately, we will also refer to the ``Red Age'' and ``Blue Age'' of each region, shown in the last two columns of Table~\ref{tab:expansion_parameters}. We estimated these using the following equation:
\begin{equation}
    t_{\rm dyn} = \frac{r_{\rm exp}}{v_{\rm exp}},
\end{equation}
\noindent where $r_{\rm exp}$ is the expansion radius and $v_{\rm exp}$ is either the redshifted or blueshifted expansion velocity. The values for $\langle t_{\rm dyn} \rangle$ (the weighted average of redshifted expansion and blueshifted expansion for each region, where the weighting is done by the percentage of PV diagrams that display an expansion signature; see Table~\ref{tab:stagnation_times}) for the ECs range from 0.04 \myr\ to 0.85 \myr. This is significantly shorter than the $1-10$ \myr\ main sequence lifetime of the OB stars that power \hii\ regions. 

One possible explanation for the observed discrepancy between the lifetime of OB stars and the dynamical ages of the regions in our sample is a negative density gradient moving away from the central ionizing source with a non-homogeneous angular density distribution, which could drive fast expansion into the surrounding, less dense medium as the \hii\ region evolves \citep{zamora2019structure}. In addition, the dynamical age of an \hii\ region begins when the region ``bursts'' out of the dense core in which the central ionizing star was born. This could be much later than when nuclear fusion is first ignited in the proto-stellar core, resulting in dynamical lifetimes much shorter than the age of the ionizing source. The low derived dynamical ages could also highlight a selection bias towards younger, brighter regions in the sample.

In addition to these factors, cloud ablation could also contribute to the large observed expansion velocities in some of the regions. As ionizing radiation erodes dense cloud surfaces, ablated material is accelerated outward, sometimes reaching velocities comparable to or exceeding the observed expansion speeds. This effect has been documented in studies such as \cite{bonne2023sofia}, which observed high-velocity \cii\ emission in RCW 79 due to photoablation of molecular clouds and radiation-driven flows. This could explain the low dynamical time estimates, as the high-velocity gas is continuously replenished while flowing outward. Consequently, \cii\ primarily traces the gas currently being expelled from the region rather than a sustained ionization front that has been expanding since the onset of the ionizing source's lifetime. Similar mechanisms may be at play in some of the regions in our sample, particularly those with strong local density gradients that result in asymmetric erosion and acceleration of material.

\subsection{Thermal Expansion vs. Wind-blown Bubbles}
\label{subsec:thermal_v_wind}

Table~\ref{tab:stagnation_times} shows an estimated theoretical range of values for $R_{s,0}$, $R_{\rm therm}$, $v_{\rm therm}$, $R_{\rm wind}$, and $v_{\rm wind}$, among other parameters. These values are estimated from Equations~\ref{eq:R_{s,0}}, ~\ref{eq:R_thermal}, ~\ref{eq:v_thermal}, ~\ref{eq:R_wind} and ~\ref{eq:v_wind}, respectively. This table also shows the observed PDR radii ($R_{\rm PDR}$) for all ECs for which values of $R_{s,0}$ could be estimated. $R_{\rm PDR}$ roughly represents the observed center of the PDR shell while $R_{\rm therm}$ represents the theoretical inner boundary, and $R_{\rm wind}$ represents the theoretical outer boundary of the PDR shell. The values are presented as pairs separated by commas that represent the minima and maxima given a range in the input parameters ($\rho_0$ = $3.35\times10^{-22}$ - $3.35\times10^{-20}$ g cm$^{-3}$ - a number density of $100-10^4$ cm$^{-3}$ multiplied by the mass of an $H_2$ molecule, $m_{H_2}=3.35*10^{-24}$; $t$ = $10^4$ yr to 1 Myr; $L_w = 10^{34} - 10^{37}$ erg s$^{-1}$). For the ECs that we are able to estimate theoretical ranges for $v_{\rm therm}$ and/or $v_{\rm wind}$ for, we compare these ranges to the PV diagram-derived expansion velocity in the region (see Section~\ref{subsec:pv_diagrams}). It is worth noting that the theoretical maxima for $v_{\rm therm}$ correspond to the earliest stages of expansion, when the swept-up shell mass is still negligible, and thus is not directly applicable to the more evolved regions considered here. In this section we will compare our observed PDR radii and derived expansion velocities to the theoretical ranges for stellar wind driven expansion and thermal pressure driven expansion in an attempt to determine the dominant mechanism of expansion in the region.

\begin{sidewaystable*}[h]
\centering
\caption{Comparison of theoretical radii and expansion velocities to observed radii and expansion velocities.} %For the theoretical calculations $n_0$ is allowed to range from 100 to $10^4$ cm$^{-3}$ with a mean particle mass of $3.35\times10^{-24}$ g (the mass of an $H_2$ molecule) and $t$ from $10^4$ yr to 10 Myr for all sources. $L_w$ values from the literature are used to estimate $R_{\rm wind}$ and $v_{\rm wind}$ for each source. Footnotes for the values in the $R_{\rm therm}$ column are references for the adopted $L_w$ values.}
\label{tab:stagnation_times}
\begin{tabular}{lccccccccc}
\hline
Region & log($L_w$) & $R_{s,0}$ [pc] & $R_{\rm therm}$ [pc] & $R_{\rm wind}$ [pc] & $R_{\rm stag}$ [pc] & $R_{\rm PDR}$ [pc] & $v_{\rm therm}$ [\kms] & $v_{\rm wind}$ [\kms] & $\langle v_{\rm exp} \rangle$ [\kms] \\
\hline
M17 & 37.00\tablenote{\footnotesize \cite{rosen2014gone}} & (0.26, 5.69) & (0.36, 13.22) & (0.39, 244.04) & (26.36, 567.92) & 2.50 & (1.70, 10.59) & (0.57, 57.03) & 16.6 \\
M42 & 35.90\tablenote{\footnotesize \cite{howarth1989stellar}} & (0.11, 2.37) & (0.20, 8.37) & (0.23, 147.26) & (10.99, 236.75) & 1.68 & (1.17, 10.39) & (0.34, 34.41) & 13.8 \\
M43 & \nodata & (0.03, 0.68) & (0.10, 4.68) & \nodata & (3.14, 67.76) & 0.13 & (0.69, 9.65) & \nodata & 3.6 \\
%N19 & \nodata & \nodata & \nodata & \nodata & \nodata & \nodata \\
%NESSIE-Aa & \nodata & \nodata & \nodata & \nodata & \nodata & \nodata \\
NGC 7538 & \nodata & (0.11, 2.37) & (0.20, 8.37) & \nodata & (10.99, 236.75) & 0.18 & (1.17, 10.39) & \nodata & 14.6 \\
RCW 120 & 35.49\tablenote{\footnotesize \cite{luisi2021stellar}} & (0.08, 1.83) & (0.17, 7.38) & (0.19, 121.64) & (8.46, 182.37) & 2.17 & (1.05, 10.29) & (0.28, 28.43) & 15.3 \\
RCW 36 & 34.00\tablenote{\footnotesize \cite{ellerbroek2013RCW36}} & (0.10, 2.05) & (0.18, 7.80) & (0.10, 61.30) & (9.5, 204.62) & 1.10 & (1.10, 10.34) & (0.14, 14.33) & 8.5 \\
RCW 79 & 36.31\tablenote{\footnotesize \cite{martins2010near}} & (0.27, 5.73) & (0.36, 13.28) & (0.29, 177.57) & (26.56, 572.29) & 10.26 & (1.71, 10.59) & (0.42, 41.50) & 12.5 \\
%W40 & \nodata & \nodata & \nodata & \nodata & \nodata & \nodata & \nodata \\
\hline
\end{tabular}  
\end{sidewaystable*}

We compare observed expansion velocities and PDR radii to theoretical expectations from both thermally and wind-driven models, using the ranges shown in Table~\ref{tab:stagnation_times}. In general, regions with observed expansion velocities significantly exceeding the maximum theoretical $v_{\rm therm}$ are likely dominated by stellar winds, while those with velocities within the range of $v_{\rm therm}$ are more consistent with thermal pressure-driven expansion.

In M17, M42, RCW 120, NGC 7538, and RCW 79, the observed expansion velocities exceed the maximum expected from thermal pressure but fall within the range predicted for stellar wind-driven shells. This suggests that stellar winds are the dominant driver of expansion in these regions. Notably, M17 and M42 each show PDR radii and expansion velocities consistent with wind-driven expansion, aligning with prior studies \citep[e.g.,][]{pabst2020expanding}. RCW 120 and RCW 79 similarly display high expansion velocities and large PDR radii indicative of stellar wind influence \citep{luisi2021stellar, bonne2023sofia}. In NGC 7538, the high expansion velocity also points to wind-driven dynamics, although the region exhibits complex internal substructure that may complicate the interpretation \citep{beuther2022feedback}.

By contrast, M43 shows a low expansion velocity consistent with thermal pressure-driven expansion, and no reliable $L_w$ estimate is available to test for wind influence. This aligns with prior work suggesting a thermally expanding shell \citep{pabst2020expanding}.

RCW 36 presents an intermediate case: both the observed radius and expansion velocity fall within the ranges predicted by both models. Spatially resolved studies reveal slower expansion in dense molecular structures and faster expansion in cavity regions, suggesting that the variation in expansion velocity is tied to density variations in the surrounding ISM \citep{bonne2022sofia}. This is consistent with the scenario proposed by \citet{bonne2022sofia}, in which the star formed in a thin molecular sheet. Once the expanding bubble broke out of the sheet, the expansion proceeded more rapidly in the perpendicular direction, appearing stellar wind driven in all directions despite asymmetry in the velocity field. This may help explain discrepancies between the stellar age and the dynamical age inferred from expansion modeling.

Overall, most regions with high expansion velocities are consistent with stellar wind-driven shells, while a smaller subset, such as M43, are better explained by thermal pressure. RCW 36 exemplifies a case where both mechanisms appear to play a role, varying spatially across the region. When studied at X-ray wavelengths, many of the rapidly expanding bubbles (such as M17, M42, RCW 36, RCW 79, and W40) reveal the presence of hot plasma generated by stellar wind shocks \citep{townsley2014massive, townsley2018massive, townsley2019massive}.

\subsection{Stagnation Time}
\label{subsec:stagnation_time}

In addition to observed and theoretical PDR radii, Table~\ref{tab:stagnation_times} also shows the estimated theoretical values for $R_{\rm stag}$ (column~4, from Equation~\ref{eq:R_stag}), and $t_{\rm stag}$ (column~7, from Equation~\ref{eq:t_stag}) for the ECs for which information is available on their ionizing sources. Comparing the theoretical $t_{\rm stag}$ value with the observationally derived $\langle t_{\rm dyn} \rangle$ value shows that the inferred dynamical time is, on average, 10\% of the expected stagnation time. For all of these calculations, it is assumed that the ambient density is $n_0 = 100$ cm$^{-3}$ and the hot and cold medium sound speeds are taken to be $c_i = 10.74$\,\kms\ and $c_o = 0.34$\,\kms\, respectively. Every EC shows signs of expansion in at least one of their PV cuts, suggesting that stagnation has not occurred in their PDRs.

\subsection{Ionizing Source vs Expansion}
\label{subsec:ionizing_source_vs_expansion}

Table~\ref{tab:ionizing_sources} shows our ECs for which information about the ionizing source(s) is available in the literature. The spectral type and approximate main sequence lifetime of the ionizing source are shown, along with the average observed expansion velocity and dynamical time. The last column shows that $t_{\rm dyn}$ is at least one order of magnitude smaller than the lifetime of the ionizing source for all ECs analyzed. If we instead use Equation~\ref{eq:t_stag} to estimate $t_{\rm dyn}$ by replacing $R_{\rm stag}$ with $R_{\rm PDR}$ (see Table~\ref{tab:stagnation_times}), we find that the dynamical times decrease by about a factor of ten.

\begin{table*}[h]
\centering
%\tiny
\caption{Ionizing Sources and Timescales}
\label{tab:ionizing_sources}
\begin{tabular}{lccccccc}
\hline
Region & Ionizing Source(s) & $F^{*}$ [photons s$^{-1}$] & $t_{\rm source}$ [\myr] & $t_{\rm stag}$ [\myr] & $\langle v_{\rm exp} \rangle$ [\kms] & $\langle t_{\rm dyn} \rangle$ [\myr] & $\frac{\langle t_{\rm source} \rangle}{t_{\rm dyn}}$ \\
\hline
M17 & O4V-O4V \tablenote{\footnotesize \cite{hanson1995identification}} & $10^{49.77}$ & $\sim\!2.5$ & 3.81 & 16.60 & 0.15 & 16.67 \\
M42 & O7V \tablenote{\footnotesize \cite{pabst2020expanding}} & $10^{48.63}$ & $\sim\!5$ & 1.59 & 13.83 & 0.12 & 41.67 \\
M43 & B0.5V \tablenote{\footnotesize \cite{pabst2020expanding}} & $10^{47.00}$ & $\sim\!10$ & 0.46 & 3.62 & 0.04 & 250.00 \\
%N19 & \nodata & \nodata & \nodata & \nodata & 8.26 & 0.47 & \nodata \\
%NESSIE-Aa & \nodata & \nodata & \nodata & \nodata & 16.40 & 0.01 & \nodata \\
NGC 7538 & O7V \tablenote{\footnotesize \cite{sandell2020NGC}} & $10^{48.63}$ & $\sim\!5$ & 1.59 & 14.57 & 0.11 & 45.45 \\
RCW 120 & O8V \tablenote{\footnotesize \cite{georgelin1970regions}} & $10^{48.29}$ & $\sim\!6$ & 1.23 & 15.33 & 0.14 & 43.48\\
RCW 36 & O8V-O9V \tablenote{\footnotesize \cite{verma1994far}} & $10^{48.44}$ & $\sim\!7$ & 1.37 & 8.47 & 0.13 & 52.63 \\
RCW 79 & O3V-O5V \tablenote{\footnotesize \cite{zavagno2006triggered}} & $10^{49.78}$ & $\sim\!2.5$ & 3.84 & 12.56 & 0.86 & 2.94 \\
%W40 & Cluster \tablenote{\footnotesize \cite{comeron2022extended}} & \nodata & \nodata & \nodata& 12.00 & 0.04 & \nodata \\
\hline
\end{tabular}  
\end{table*}

\subsubsection{$F^*$ and $L_w$ vs Expansion Velocity}
\label{subsubsec:F*_L_w_vs_exp_vel}

We examine whether the expansion velocities of the \hii\ region PDR fronts correlates with either the ionizing UV flux ($F^*$) or the mechanical luminosity ($L_w$) of the central source(s).

For ionizing photon flux, we find no clear trend: regions with high UV output, such as M17 and RCW 79, show a wide range of expansion velocities from approximately 13 to 17.5 \kms. Other regions with similar or slightly lower fluxes, like RCW 120 and M42, exhibit comparable expansion rates. This spread does not agree with the prediction of Equation~\ref{eq:v_thermal} that the expansion rate to scale with ionizing flux due to increased thermal pressure.

Several regions with known ionizing fluxes show no detectable expansion. These are typically not bubble-shaped and fall near or below our estimated detection threshold of $\sim\!3$ \kms. This suggests that morphology, orientation, and local environmental conditions influence whether we observe expansion. All regions with expansion signatures in 100\% of their PV cuts appear as bubbles in \cii\ emission, supporting the idea that symmetrical regions are more likely to exhibit clear expansion.

We also compare expansion velocity with mechanical luminosity ($L_w$) where available and again find no consistent trend. M17, which has the highest $L_w$ ($\sim\!10^{37}$ erg s$^{-1}$), expands rapidly, but RCW 79, with a similar $L_w$, expands more slowly. Conversely, RCW 36 shows moderate expansion ($\sim\!8-9$ \kms) despite having one of the lowest mechanical luminosities in the sample ($\sim\!10^{34-35}$ erg s$^{-1}$). These results indicate that while stellar winds may contribute to expansion, they do not dominate. Expansion velocity likely depends on a combination of factors, including UV flux, mechanical feedback, density structure, morphology, and projection effects.

\subsubsection{Ionizing Source Lifetime vs Dynamical Ages}
\label{subsubsec:source_life_v_dyn_age}

When comparing the expansion-derived dynamical ages of our ECs to the estimated main-sequence lifetimes of their ionizing sources, in every case, the dynamical age falls well below the stellar lifetime—typically by at least an order of magnitude (and in some cases by a factor of 100 or more.) This aligns well with results from other studies where dynamical ages estimated from classical expansion models are consistently shorter than the actual lifespans of the stars driving the ionization.

This discrepancy in age reflects the limitations of idealized models that do not account for all mechanisms at play. Several factors can delay or suppress the observable expansion of an \hii\ region, including confinement by dense natal material, time-variable ionizing flux \citep{galvan2011time}, or expansion into a clumpy medium. For example, the ionization front may remain trapped within the dense core for tens to hundreds of thousands of years before breaking out and driving large-scale expansion \citep{keto2002evolution}. During this time, the star continues to age, but the region’s expansion ``clock'' has not yet started. Once the front breaks out, the expansion begins rapidly, but the inferred dynamical age—based on the current radius and expansion velocity—only captures the post-breakout phase.

% We therefore expect dynamical ages to be systematically smaller than stellar lifetimes, and our observations support that expectation. The magnitude of this offset varies across the sample, but the trend is consistent: no region in our analysis shows a dynamical age exceeding the lifetime of its ionizing source.}

\subsection{Individual Regions}
\label{subsec:individual_regions}

In this section we discuss the results for three of our ECs that have not yet been discussed in the literature.

\subsubsection{G083+936}
\label{subsubsec:G083+936}

\hii\ region G083+936 appears as a compact bubble in \cii\ emission with an opening in the southwest of its PDR, as shown in the upper left panel of Figure~\ref{fig:G083_master}. The PV analysis for this source shows that all 16 of the PV diagram axes explored display both redshifted and blueshifted expansion signatures. One of the PV diagrams for this regions is shown in Figure~\ref{fig:G083P936_PV}. This is the first \hii\ region for which both redshifted and blueshifted expansion is present along all probed axes (see Table~\ref{tab:ionizing_sources}). The $0.09 \pm 0.03$\,pc observed PDR radius of this source is the the smallest in the sample (sample average: $\sim\!2.48$\,pc). The observed blueshifted expansion in the source is $-8.86 \pm 0.47$\,\kms, which is lower than the $\sim\!-11.97$\,\kms\ sample average. The observed redshifted expansion in the source is $12.61 \pm 0.57$\,\kms, which is also slightly lower than the $\sim\!13.18$\,\kms\ sample average. The redshifted and blueshifted dynamical ages for the source are also the smallest in the sample. This together with the small observed PDR radius suggests that G083+936 may be relatively young and less evolved compared to to other \hii\ regions in the sample.

\begin{figure*}[ht]
    \centering
    \includegraphics[width=\textwidth]{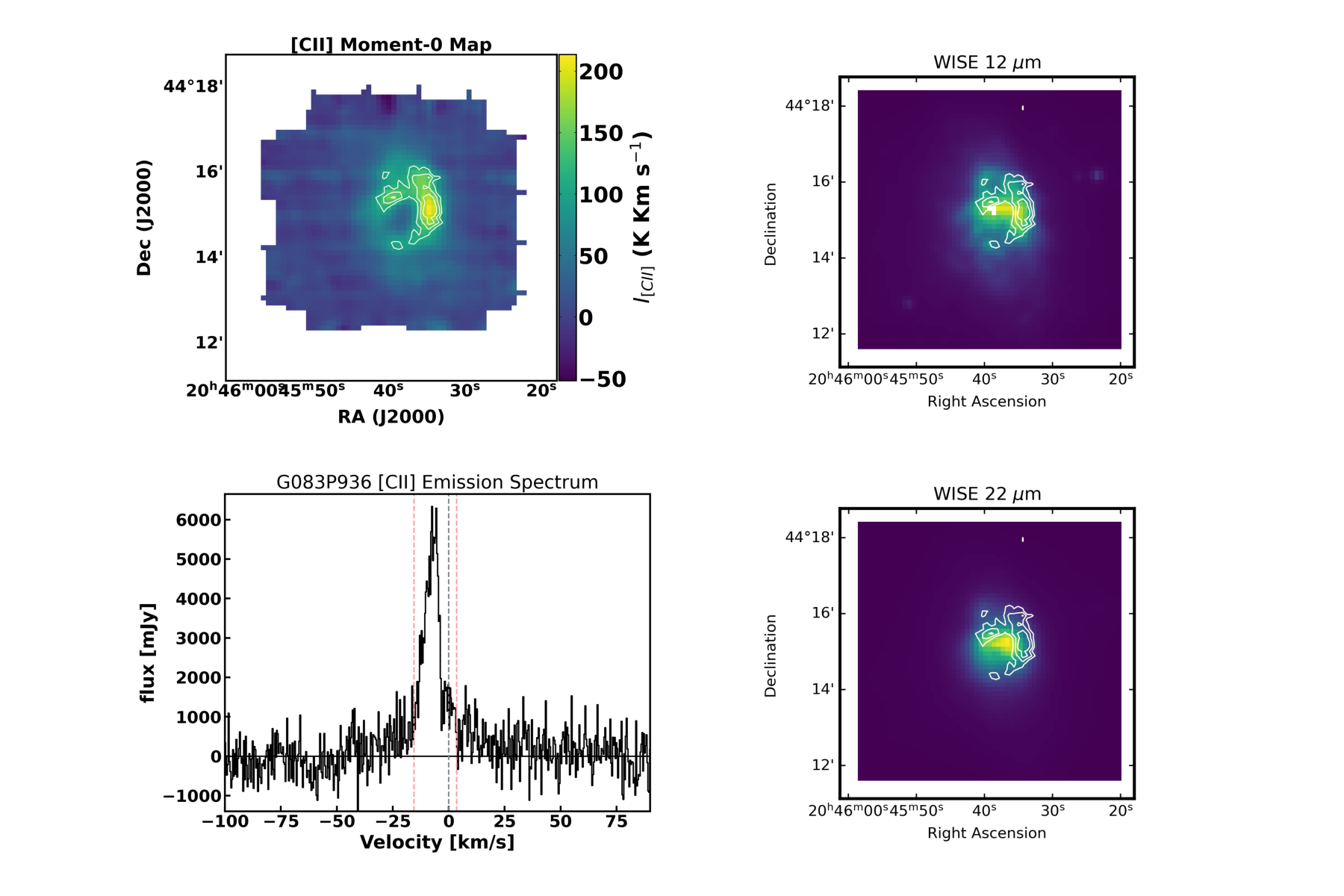}
    \caption{\textit{Top Left:} Moment-0 map of SOFIA \cii\ emission for \hii\ region G083+936. Overlaid contours show 4$\sigma$, 5$\sigma$, 7$\sigma$, 9$\sigma$, 12$\sigma$, 15$\sigma$, and 18$\sigma$ levels of detection for \cii\ emission. \textit{Top Right:}  WISE 12 \micron\ emission with \cii\ contours overlaid. \textit{Bottom Left:} SOFIA \cii\ spectrum of entire field. \textit{Bottom Right:} WISE 22 \micron\ emission with \cii\ contours overlaid. }
    \label{fig:G083_master}
\end{figure*}

\begin{figure*}[ht]
    \centering
    \includegraphics[width=\textwidth]{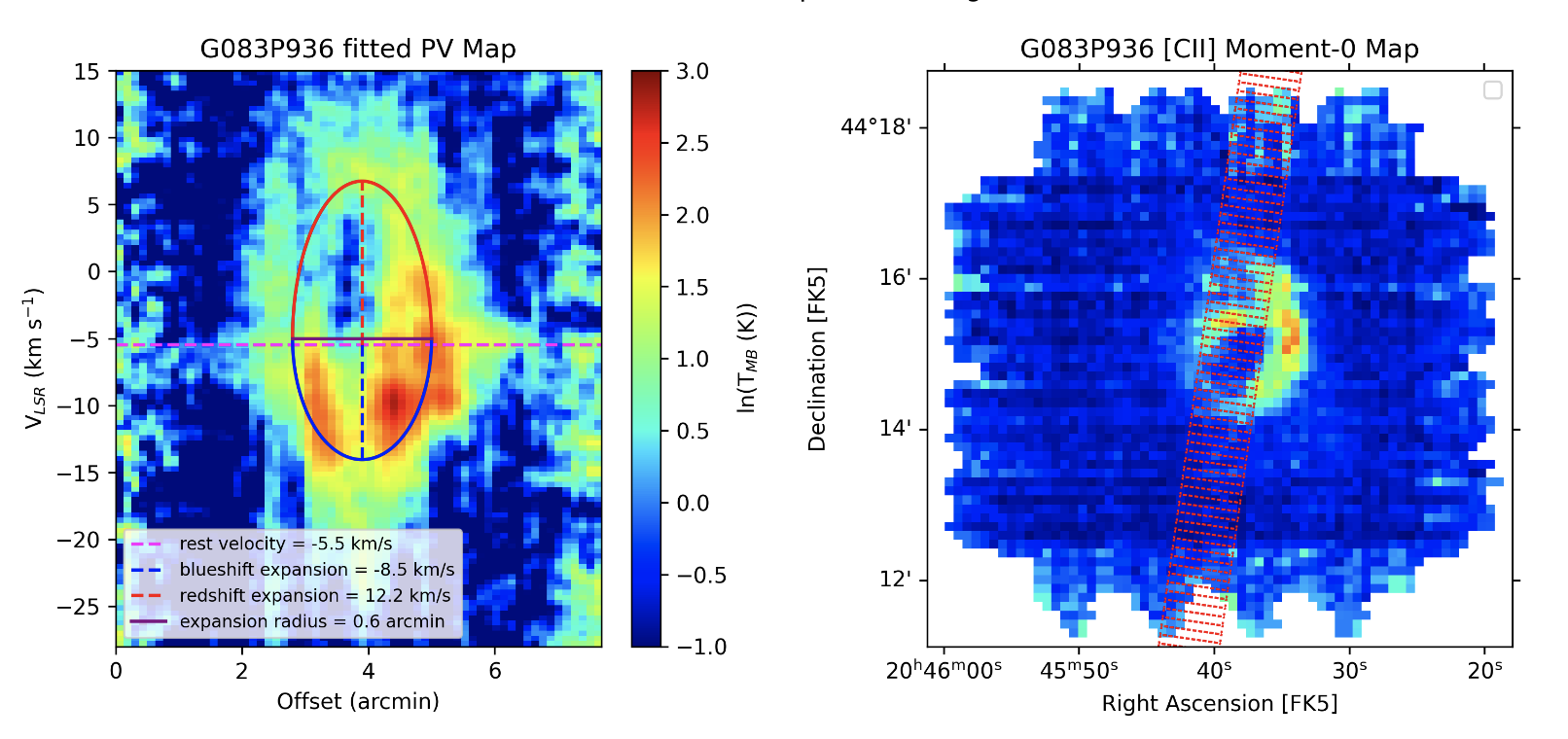}
    \caption{\textit{Left:} PV diagram of G083+936. \textit{Right:} \cii\ moment-0 map of G083+936. The red are cutting across the field is the area from which the PV diagram on the left is derived.}
    \label{fig:G083P936_PV}
\end{figure*}

\subsubsection{G316+796}
\label{subsubsec:G316+796}

\hii\ region G316+796 appears as a diffuse bubble in \cii\ emission with openings in the northeast, east, and southwest of its PDR, as shown in the upper left panel of Figure~\ref{fig:G316_master}. The PV analysis for this source shows that eight of the 16 PV diagram axes explored display redshifted expansion signatures, while none display any blueshifted expansion signatures. The non-detection of blue- or redshifted emission in the PV diagrams of any region may therefore be a sensitivity limitation rather than definitive evidence of its absence. One of the PV diagrams for this regions is shown in Figure~\ref{fig:G316P796_PV}. The lack of a blueshifted counterpart suggests that the observed kinematics may be influenced by a line-of-sight orientation effect or an inherent asymmetry in the expansion process. The $2.05 \pm 0.23$\,pc observed PDR radius of this source is slightly below the sample average of $\sim\!2.48$\,pc. The observed redshifted expansion in the source is $12.95 \pm 0.36$\,\kms, which is within range of the $\sim\!13.18$\,\kms\ sample average. The dynamical ages for the source are also among the smallest in the sample. The redshifted dynamical age for the source is $0.16\pm0.02$ Myr, nearly in range of the average redshifted age in the sample of $\sim\!0.21$ Myr.

\begin{figure*}[ht]
    \centering
    \includegraphics[width=\textwidth]{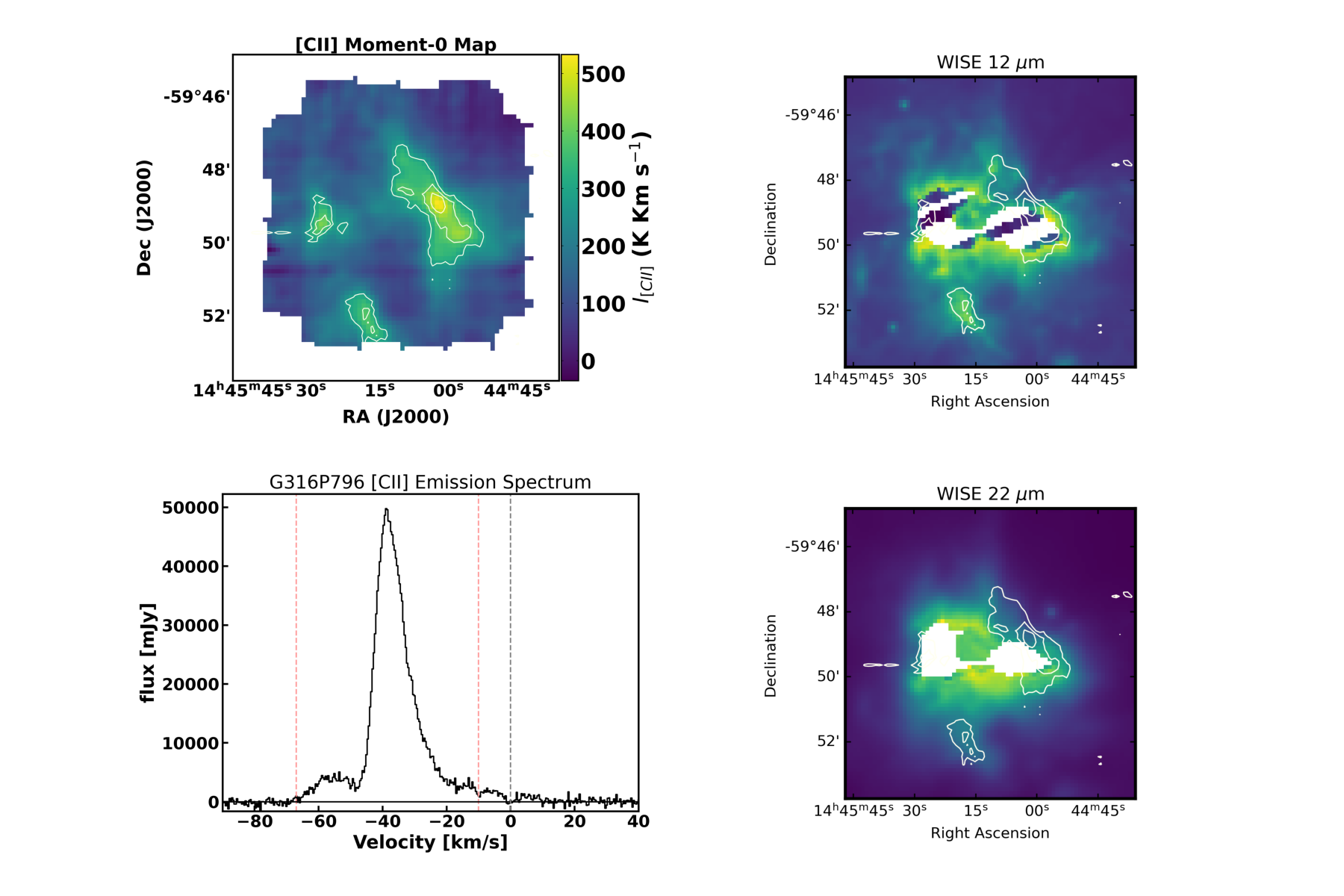}
    \caption{\textit{Top Left:} Moment-0 map of SOFIA \cii\ emission for \hii\ region G316+796. Overlaid contours show 4$\sigma$, 5$\sigma$, 7$\sigma$, 9$\sigma$, 12$\sigma$, 15$\sigma$, and 18$\sigma$ levels of detection for \cii\ emission. \textit{Top Right:}  WISE 12 \micron\ emission (saturated) with \cii\ contours overlaid. \textit{Bottom Left:} SOFIA \cii\ spectrum of entire field. \textit{Bottom Right:} WISE 22 \micron\ emission (saturated) with \cii\ contours overlaid.}
    \label{fig:G316_master}
\end{figure*}

\begin{figure*}[ht]
    \centering
    \includegraphics[width=\textwidth]{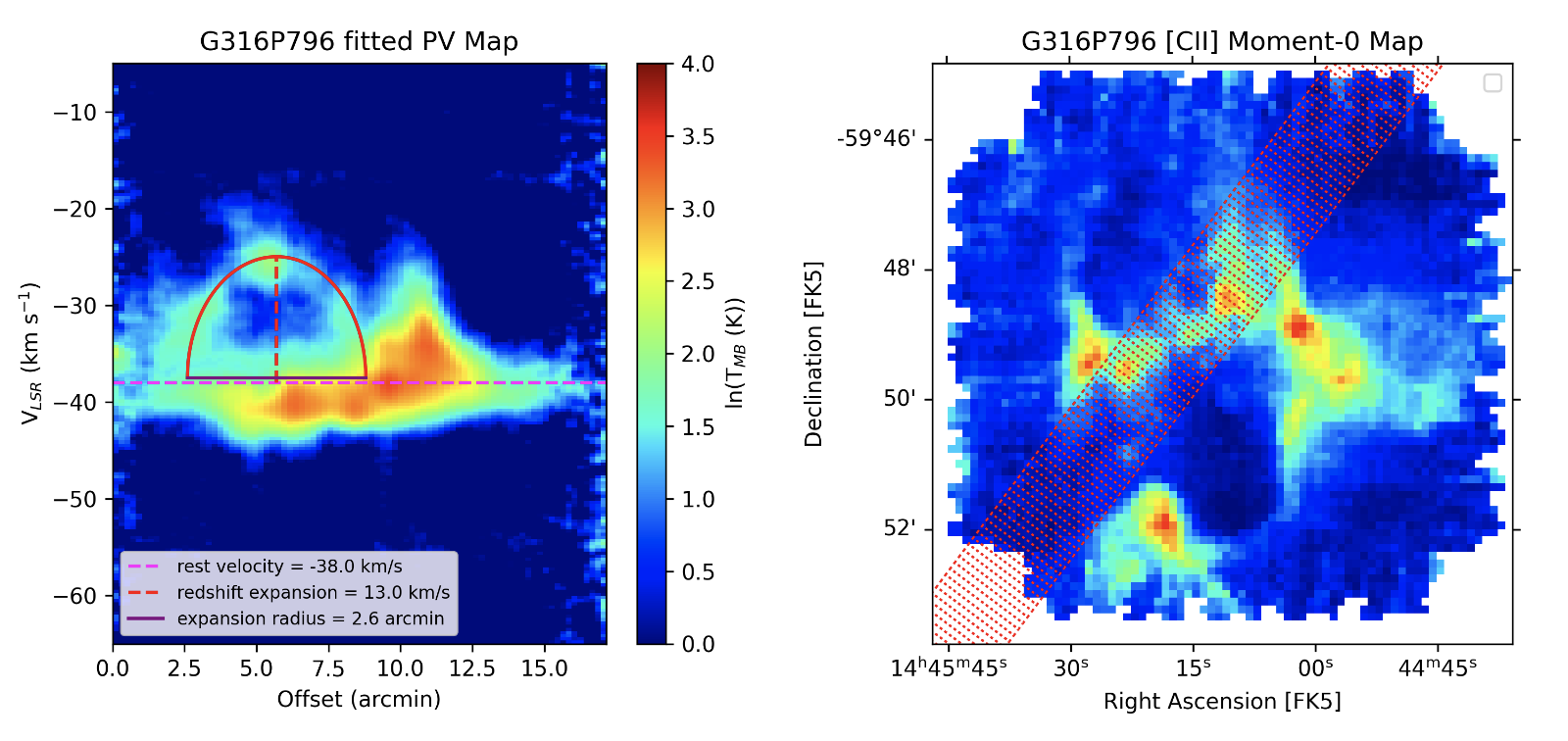}
    \caption{\textit{Left:} PV diagram of G316+796. \textit{Right:} \cii\ moment-0 map of G316+796. The red are cutting across the field is the area from which the PV diagram on the left is derived.}
    \label{fig:G316P796_PV}
\end{figure*}

\subsubsection{NESSIE-A Sub-Bubble (NESSIE-Aa)}
\label{subsubsec:NESSIE-Aa}

The \hii\ region NESSIE-A appears as a long, wavy filament of ionized gas, as shown in the top left panel of Figure~\ref{fig:NESSIE-A_master}. The filamentary portion of the region itself does not display signs of expansion; however, the sub-bubble in NESSIE-A (outlined by the red circle) exhibits distinct morphological and kinematic properties within the larger region. Denoted as NESSIE-Aa, the sub-bubble has a compact observed radius of 0.18 pc (the smallest in the sample of ECs) and displays exclusively blueshifted expansion with a velocity of -16.4\,\kms\ in two of the 16 PV diagrams analyzed. This velocity is among the highest observed in the sample. The lack of a redshifted counterpart suggests that the observed kinematics may be influenced by a line-of-sight orientation effect or an inherent asymmetry in the expansion process. The compact size and high expansion velocity of NESSIE-Aa may indicate that the sub-bubble is in an early phase of evolution. The proximity of NESSIE-Aa to the main structure of NESSIE-A suggests potential interactions between the two, where the sub-bubble could be a site of new and ongoing triggered star formation, as suggested in \cite{jackson2024absorption}.

\begin{figure*}[ht]
    \centering
    \includegraphics[width=\textwidth]{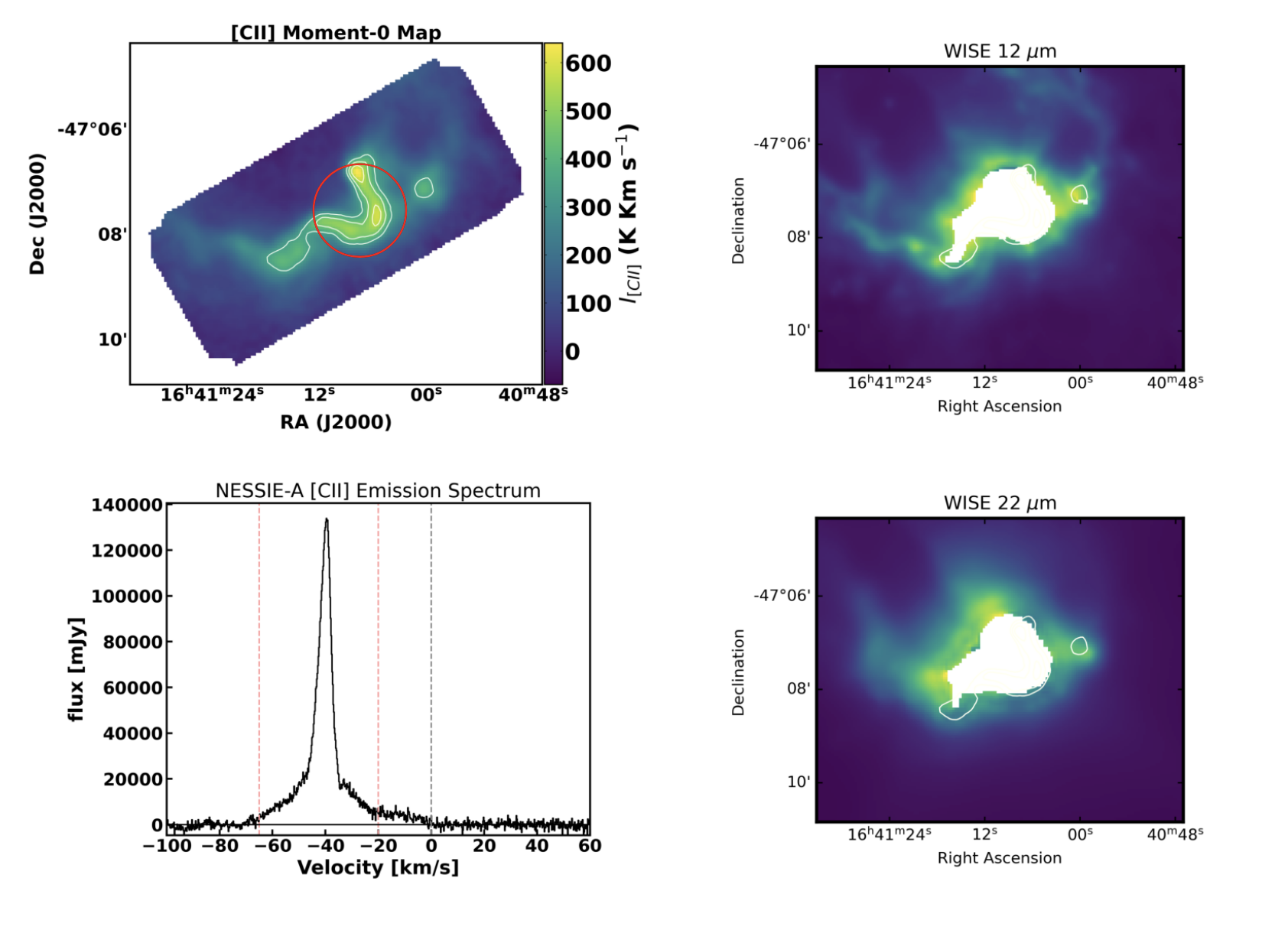}
    \caption{\textit{Top Left:} Moment-0 map of SOFIA \cii\ emission with 4$\sigma$, 5$\sigma$, 7$\sigma$, 9$\sigma$, 12$\sigma$, 15$\sigma$, and 18$\sigma$ \cii\ emission contours overlaid. The red circle indicates the sub-bubble NESSIE-Aa. \textit{Top Right:} WISE 12 \micron\ emission (saturated) with \cii\ contours overlaid. \textit{Bottom Left:} SOFIA \cii\ spectrum of entire field. \textit{Bottom Right:} WISE 22 \micron\ emission (saturated) with \cii\ contours overlaid.}
    \label{fig:NESSIE-A_master}
\end{figure*}

\begin{figure*}[ht]
    \centering
    \includegraphics[width=\textwidth]{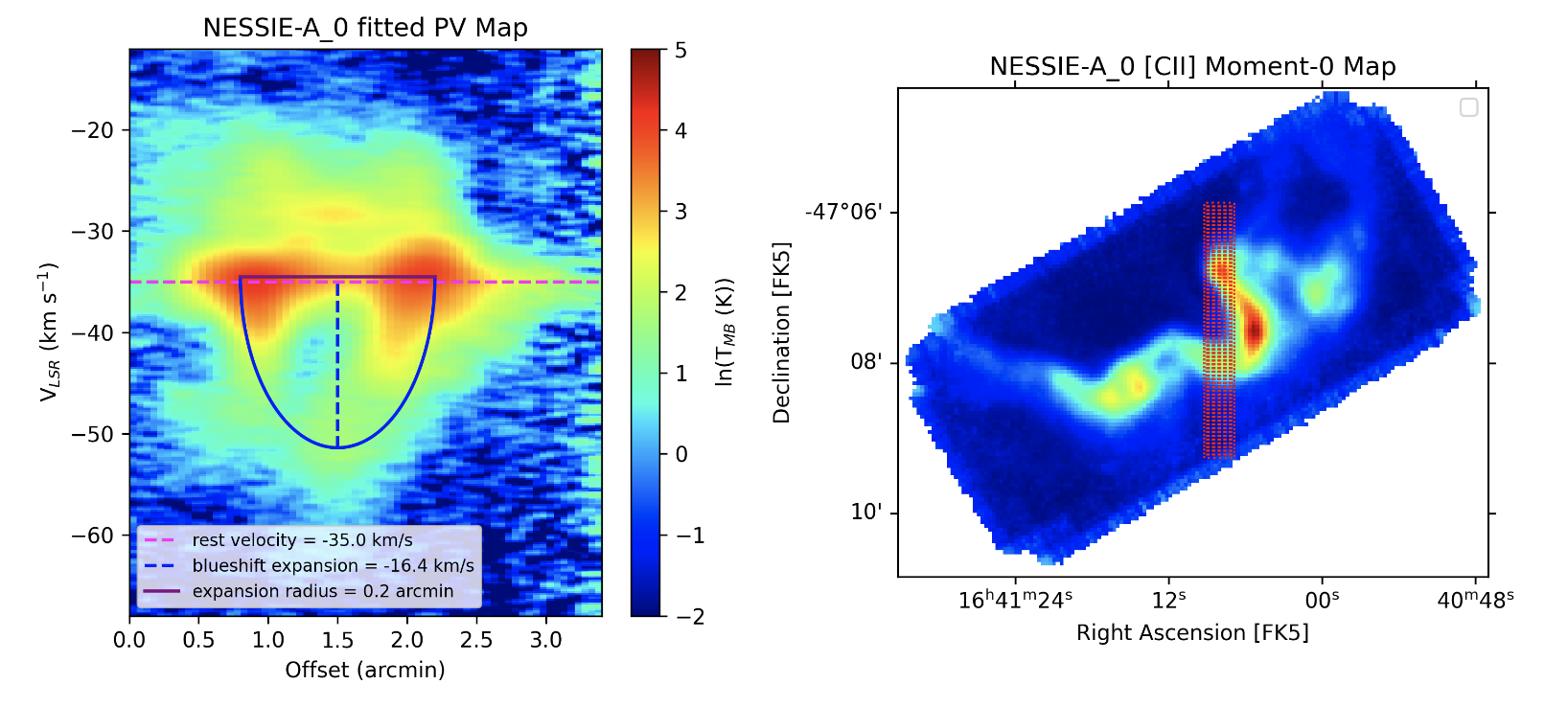}
    \caption{\textit{Left:} PV diagram of the sub-bubble in NESSIE-A. \textit{Right:} \cii\ moment-0 map of NESSIE-A. The red area cutting across the field is the area from which the PV diagram on the left is derived.}
    \label{fig:NESSIE-A_PV}
\end{figure*}

\section{Conclusion}
\label{sec:conclusion}

We conducted a kinematic analysis of 35 \hii\ regions mapped in \cii\ 158 \micron\ emission by SOFIA to investigate the expansion of their surrounding PDRs. Our key findings are summarized below:

\begin{itemize}
    \item PV Detection of Expansion:
    We use PV diagrams to identify expansion signatures in 12 of the 35 regions studied ($\sim$34\%). Most expansion appears blueshifted, while a smaller fraction shows redshifted or both blueshifted and redshifted. The average expansion velocity among our Expansion Candidates (ECs) is 12.21\,\kms, supporting the idea that \hii\ regions actively expand due to feedback from their ionizing stars.

    \item Regions without PV-detected Expansion: Of the 23 regions without detected expansion, all have radii smaller than their predicted stagnation radii (where $F^*$ is available), suggesting they have not yet stagnated and are likely still expanding at undetectable or asymmetric rates.
    
    \item Velocity Residual Maps: 
    Velocity residual maps support the PV analysis and also allow us to examine deviations from homogeneous expansion. Only one of the ten residual maps falls within the error bars of its corresponding PV diagram finding, while the others deviate by only a few \kms\ but still yield consistent morphological fits. Simulations show that higher expansion velocities and S/N values yield more reliable expansion velocity fits.

    \item Expansion Mechanisms:
    Comparisons of observed expansion velocities with theoretical models suggest different driving mechanisms. Thermal pressure appears to dominate in M43, while stellar wind pressure is the likely driver in M17, M42, RCW 120, and RCW 79. The expansion in RCW 36 falls within both theoretical models, making it difficult to determine if one mechanism dominates over the other.

    \item Radial Expansion Asymmetry:
    Some regions exhibit significant variations in expansion velocity along different axes, likely influenced by local density variations. Differences in blueshifted and redshifted expansion velocities may reflect foreground and background density differences.

    \item Dynamical Ages vs. Stellar Lifetimes: 
    Estimated dynamical ages of ECs are consistently an order of magnitude or more shorter than the main-sequence lifetimes of their ionizing sources. Possible explanations include delayed onset of expansion due to confinement by dense natal material, expansion into a clumpy or porous medium, and projection effects. High observed expansion velocities may also result from photoablation of molecular clouds along the ionization front.

\end{itemize}

The expansion and feedback of \hii\ regions play a crucial role in shaping the ISM, regulating star formation rates, and influencing galactic evolution. Future studies of \hii\ region expansion should focus on increasing the sample size, which will help refine our understanding of how \hii\ regions evolve over time. More detailed constraints on ionizing sources will also help to disentangle the mechanisms driving expansion.

\newpage
%\section{Appendix (Supplementary Materials)}
%\label{sec:appendix}

%\subsubsection{WISE Emission}
%\label{subsec:wise_emission}
%We observe and stack (when not saturated) WISE emission from four bands (3.4, 4.6, 12 and 22 \micron).

%Each of these wavelengths plays an important role in understanding the properties of \hii\ regions. Observed emission at 3.4 and 4.6 \micron\ is associated with the presence of stars.

%At a wavelength of 12 \micron\, observed emission is dominated by continuum emission from warm dust \citep{draine2007infrared}. This is the most useful of the WISE bands when tracing the PDR of an \hii\ region. 12 \micron\ emission is also associated with the presence of polycyclic aromatic hydrocarbons (PAHs) as well as dust continuum \citep{henning1998carbon}. PAHs are organic molecules that are excited by UV radiation from OB stars in \hii\ regions \citep{allamandola1989interstellar}.

%Observed emission at 22 \micron\ originates from cooler dust components, tracing the presence of cooler dust grains in the \hii\ regions \citep{draine2007infrared}. By overlaying contours of \cii\ moment-0 maps onto the stacked WISE photometric observations at these different wavelengths, a more complete picture of the relationship between the \cii\ emission and the distribution of dust and organic molecules within the \hii\ regions can be achieved. Some PAHs are also detectable around this wavelength.

\begin{acknowledgments}
\begin{centering}
{\sc Acknowledgments}\\
\end{centering}
This study is based on observations made with the NASA/DLR Stratospheric Observatory for Infrared Astronomy (SOFIA). SOFIA is jointly operated by the Universities Space Research Association Inc. (USRA), under NASA contract NNA17BF53C, and the Deutsches SOFIA Institut (DSI), under DLR contract 50 OK 0901 to the University of Stuttgart. upGREAT is a development by the MPIfR and the KOSMA/Universit\"at zu K\"oln, in cooperation with the DLR Institut f\"ur Optische Sensorsysteme.\\
TF and LDA acknowledge support from Universities Space Research Association grant ``\hii\ Region Dynamics Revealed by \cii\ Emission'' \#09-0520. N.S. acknowledges support (FEEDBACK-plus project) by the BMWI via DLR, Projekt Number 50OR2217.  This work was supported by the CRC1601 (SFB 1601 sub-projects A6, B2) funded by the DFG (German Research Foundation) – 500700252. \\ TF and LDA would also like to thank Dr. James Jackson, the former director of the NSF Green Bank Observatory, and Dr. Robin JR Williams for helping with specific research questions.

\end{acknowledgments}

\bibliographystyle{aasjournal}
\bibliography{bib}

\begin{thebibliography}{}
\expandafter\ifx\csname natexlab\endcsname\relax\def\natexlab#1{#1}\fi
\providecommand{\url}[1]{\href{#1}{#1}}
\providecommand{\dodoi}[1]{doi:~\href{http://doi.org/#1}{\nolinkurl{#1}}}
\providecommand{\doeprint}[1]{\href{http://ascl.net/#1}{\nolinkurl{http://ascl.net/#1}}}
\providecommand{\doarXiv}[1]{\href{https://arxiv.org/abs/#1}{\nolinkurl{https://arxiv.org/abs/#1}}}

\bibitem[{Anderson {et~al.}(2010)Anderson, Zavagno, Rod{\'o}n, Russeil, Abergel, Ade, Andr{\'e}, Arab, Baluteau, Bernard, {et~al.}}]{anderson2010physical}
Anderson, L., Zavagno, A., Rod{\'o}n, J., {et~al.} 2010, Astronomy \& Astrophysics, 518, L99

\bibitem[{Anderson {et~al.}(2012)Anderson, Zavagno, Deharveng, Abergel, Motte, Andr{\'e}, Bernard, Bontemps, Hennemann, Hill, {et~al.}}]{anderson2012dust}
Anderson, L., Zavagno, A., Deharveng, L., {et~al.} 2012, Astronomy \& Astrophysics, 542, A10

\bibitem[{Anderson {et~al.}(2019)Anderson, Makai, Luisi, Andersen, Russeil, Samal, Schneider, Tremblin, Zavagno, Kirsanova, {et~al.}}]{anderson2019origin}
Anderson, L., Makai, Z., Luisi, M., {et~al.} 2019, The Astrophysical Journal, 882, 11

\bibitem[{Anderson {et~al.}(2014)Anderson, Bania, Balser, Cunningham, Wenger, Johnstone, \& Armentrout}]{anderson2014wise}
Anderson, L.~D., Bania, T., Balser, D.~S., {et~al.} 2014, The Astrophysical Journal Supplement Series, 212, 1

\bibitem[{Balser {et~al.}(2001)Balser, Goss, \& De~Pree}]{balser2001vla}
Balser, D.~S., Goss, W., \& De~Pree, C. 2001, The Astronomical Journal, 121, 371

\bibitem[{Bania {et~al.}(1997)Bania, Balser, Rood, Wilson, \& Wilson}]{bania19973he}
Bania, T., Balser, D.~S., Rood, R.~T., Wilson, T., \& Wilson, T. 1997, The Astrophysical Journal Supplement Series, 113, 353

\bibitem[{Barman {et~al.}(2022)Barman, Neelamkodan, Madden, Sewilo, Kemper, Tokuda, Sanyal, \& Onishi}]{barman2022study}
Barman, S., Neelamkodan, N., Madden, S.~C., {et~al.} 2022, The Astrophysical Journal, 930, 100

\bibitem[{Belloni \& Mereghetti(1994)}]{belloni1994rosat}
Belloni, T., \& Mereghetti, S. 1994, Astronomy and Astrophysics, Vol. 286, p. 935-942 (1994), 286, 935

\bibitem[{Benaglia {et~al.}(2013)Benaglia, Koribalski, Peri, Marti, Sanchez-Sutil, Dougherty, \& Noriega-Crespo}]{benaglia2013high}
Benaglia, P., Koribalski, B., Peri, C.~S., {et~al.} 2013, Astronomy \& Astrophysics, 559, A31

\bibitem[{Bertoldi(1989)}]{bertoldi1989photoevaporation}
Bertoldi, F. 1989, Astrophysical Journal, Part 1 (ISSN 0004-637X), vol. 346, Nov. 15, 1989, p. 735-755. Research sponsored by NASA., 346, 735

\bibitem[{Beuther {et~al.}(2022)Beuther, Schneider, Simon, Suri, Ossenkopf-Okada, Kabanovic, R{\"o}llig, Guevara, Tielens, Sandell, {et~al.}}]{beuther2022feedback}
Beuther, H., Schneider, N., Simon, R., {et~al.} 2022, Astronomy \& Astrophysics, 659, A77

\bibitem[{Bisbas {et~al.}(2015)Bisbas, Haworth, Williams, Mackey, Tremblin, Raga, Arthur, Baczynski, Dale, Frostholm, {et~al.}}]{bisbas2015starbench}
Bisbas, T.~G., Haworth, T., Williams, R., {et~al.} 2015, Monthly Notices of the Royal Astronomical Society, 453, 1324

\bibitem[{Bonne {et~al.}(2022)Bonne, Schneider, Garc{\'\i}a, Bij, Broos, Fissel, Guesten, Jackson, Simon, Townsley, {et~al.}}]{bonne2022sofia}
Bonne, L., Schneider, N., Garc{\'\i}a, P., {et~al.} 2022, The Astrophysical Journal, 935, 171

\bibitem[{Bonne {et~al.}(2023{\natexlab{a}})Bonne, Kabanovic, Schneider, Zavagno, Keilmann, Simon, Buchbender, G{\"u}sten, Jacob, Jacobs, {et~al.}}]{bonne2023sofia}
Bonne, L., Kabanovic, S., Schneider, N., {et~al.} 2023{\natexlab{a}}, Astronomy \& Astrophysics, 679, L5

\bibitem[{Bonne {et~al.}(2023{\natexlab{b}})Bonne, Bontemps, Schneider, Simon, Clarke, Csengeri, Chambers, Graf, Jackson, Klein, {et~al.}}]{bonne2023unveiling}
Bonne, L., Bontemps, S., Schneider, N., {et~al.} 2023{\natexlab{b}}, The Astrophysical Journal, 951, 39

\bibitem[{Cambr{\'e}sy {et~al.}(2013)Cambr{\'e}sy, Marton, Feher, T{\'o}th, \& Schneider}]{cambresy2013young}
Cambr{\'e}sy, L., Marton, G., Feher, O., T{\'o}th, L.~V., \& Schneider, N. 2013, Astronomy \& Astrophysics, 557, A29

\bibitem[{Carral {et~al.}(2002)Carral, Kurtz, Rodr{\'\i}guez, Menten, Cant{\'o}, \& Arceo}]{carral2002detection}
Carral, P., Kurtz, S.~E., Rodr{\'\i}guez, L.~F., {et~al.} 2002, The Astronomical Journal, 123, 2574

\bibitem[{Churchwell {et~al.}(2006)Churchwell, Povich, Allen, \& et~al.}]{churchwell2006bubbling}
Churchwell, E., Povich, M.~S., Allen, D., \& et~al. 2006, The Astrophysical Journal, 649, 759

\bibitem[{Comer{\'o}n {et~al.}(2022)Comer{\'o}n, Djupvik, \& Schneider}]{comeron2022extended}
Comer{\'o}n, F., Djupvik, A., \& Schneider, N. 2022, Astronomy \& Astrophysics, 665, A76

\bibitem[{Dale {et~al.}(2015)Dale, Haworth, \& Bressert}]{dale2015dangers}
Dale, J., Haworth, T., \& Bressert, E. 2015, Monthly Notices of the Royal Astronomical Society, 450, 1199

\bibitem[{Damiani {et~al.}(2019)Damiani, Prisinzano, Micela, \& Sciortino}]{damiani2019wide}
Damiani, F., Prisinzano, L., Micela, G., \& Sciortino, S. 2019, Astronomy \& Astrophysics, 623, A25

\bibitem[{Deharveng \& Zavagno(2010)}]{deharveng2010observations}
Deharveng, L., \& Zavagno, A. 2010, Proceedings of the International Astronomical Union, 6, 239

\bibitem[{Draine(2011)}]{Draine2011}
Draine, B.~T. 2011, Physics of the Interstellar and Intergalactic Medium (Princeton University Press)

\bibitem[{Ellerbroek {et~al.}(2013)Ellerbroek, Bik, Kaper, Maaskant, Paalvast, Tramper, Sana, Waters, \& Balog}]{ellerbroek2013RCW36}
Ellerbroek, L., Bik, A., Kaper, L., {et~al.} 2013, Astronomy \& Astrophysics, 558, A102

\bibitem[{Elmegreen \& Lada(1977)}]{elmegreen1977sequential}
Elmegreen, B.~G., \& Lada, C.~J. 1977, Astrophysical Journal, Part 1, vol. 214, June 15, 1977, p. 725-741., 214, 725

\bibitem[{Figueira {et~al.}(2017)Figueira, Zavagno, Deharveng, Russeil, Anderson, Men’shchikov, Schneider, Hill, Motte, M{\`e}ge, {et~al.}}]{figueira2017star}
Figueira, M., Zavagno, A., Deharveng, L., {et~al.} 2017, Astronomy \& Astrophysics, 600, A93

\bibitem[{Fleener {et~al.}(2009)Fleener, Payne, Chu, Chen, \& Gruendl}]{fleener2009massive}
Fleener, C.~E., Payne, J.~T., Chu, Y.-H., Chen, C.-H.~R., \& Gruendl, R.~A. 2009, The Astronomical Journal, 139, 158

\bibitem[{Franco {et~al.}(2000)Franco, Kurtz, Garc{\'\i}a-Segura, \& Hofner}]{franco2000evolution}
Franco, J., Kurtz, S., Garc{\'\i}a-Segura, G., \& Hofner, P. 2000, Astrophysics and Space Science, 272, 169

\bibitem[{Franco {et~al.}(1989)Franco, Tenorio-Tagle, \& Bodenheimer}]{franco1989expansion}
Franco, J., Tenorio-Tagle, G., \& Bodenheimer, P. 1989, in International Astronomical Union Colloquium, Vol. 120, Cambridge University Press, 96--103

\bibitem[{Galv{\'a}n-Madrid {et~al.}(2011)Galv{\'a}n-Madrid, Peters, Keto, Mac~Low, Banerjee, \& Klessen}]{galvan2011time}
Galv{\'a}n-Madrid, R., Peters, T., Keto, E.~R., {et~al.} 2011, Monthly Notices of the Royal Astronomical Society, 416, 1033

\bibitem[{Garc{\'\i}a-Rojas {et~al.}(2006)Garc{\'\i}a-Rojas, Esteban, Peimbert, Costado, Rodr{\'\i}guez, Peimbert, \& Ruiz}]{garcia2006faint}
Garc{\'\i}a-Rojas, J., Esteban, C., Peimbert, M., {et~al.} 2006, Monthly Notices of the Royal Astronomical Society, 368, 253

\bibitem[{Geist {et~al.}(2022)Geist, Gallagher, Kotulla, Oskinova, Hamann, Ramachandran, Sabbi, Smith, Kniazev, Nota, {et~al.}}]{geist2022ionization}
Geist, E., Gallagher, J., Kotulla, R., {et~al.} 2022, Publications of the Astronomical Society of the Pacific, 134, 064301

\bibitem[{Georgelin \& Georgelin(1970)}]{georgelin1970regions}
Georgelin, Y., \& Georgelin, Y. 1970, Astronomy \& Astrophysics Supplement Series Vol. 3, No. 1, p. 1, 3, 1

\bibitem[{Goodman {et~al.}(2014)Goodman, Alves, Beaumont, Benjamin, Borkin, Burkert, Dame, Jackson, Kauffmann, Robitaille, {et~al.}}]{goodman2014bones}
Goodman, A.~A., Alves, J., Beaumont, C.~N., {et~al.} 2014, The Astrophysical Journal, 797, 53

\bibitem[{Graczyk {et~al.}(2020)Graczyk, Pietrzy{\'n}ski, Thompson, Gieren, Zgirski, Villanova, Gorski, Wielgorski, Karczmarek, Narloch, {et~al.}}]{graczyk2020distance}
Graczyk, D., Pietrzy{\'n}ski, G., Thompson, I.~B., {et~al.} 2020, The Astrophysical Journal, 904, 13

\bibitem[{Guevara {et~al.}(2020)Guevara, Stutzki, Ossenkopf-Okada, Simon, P{\'e}rez-Beaupuits, Beuther, Bihr, Higgins, Graf, \& G{\"u}sten}]{guevara2020c}
Guevara, C., Stutzki, J., Ossenkopf-Okada, V., {et~al.} 2020, Astronomy \& Astrophysics, 636, A16

\bibitem[{Güsten {et~al.}(2007)Güsten, Heyminck, Wiesemeyer, Groppi, de~Breuck, Siringo, \& Zinnecker}]{Guesten2014}
Güsten, R., Heyminck, S., Wiesemeyer, H., {et~al.} 2007, Astronomy \& Astrophysics, 454, L13

\bibitem[{Hanson \& Conti(1995)}]{hanson1995identification}
Hanson, M.~M., \& Conti, P.~S. 1995, The Astrophysical Journal, 448, L45

\bibitem[{Heyminck {et~al.}(2012)Heyminck, Graf, G{\"u}sten, Stutzki, H{\"u}bers, \& Hartogh}]{heyminck2012great}
Heyminck, S., Graf, U., G{\"u}sten, R., {et~al.} 2012, Astronomy \& Astrophysics, 542, L1

\bibitem[{Hollenbach \& Tielens(1999)}]{hollenbach1999photodissociation}
Hollenbach, D.~J., \& Tielens, A. 1999, Reviews of Modern Physics, 71, 173

\bibitem[{Howarth \& Prinja(1989)}]{howarth1989stellar}
Howarth, I.~D., \& Prinja, R.~K. 1989, Astrophysical Journal Supplement Series (ISSN 0067-0049), vol. 69, March 1989, p. 527-592., 69, 527

\bibitem[{Jackson \& Kraemer(1999)}]{jackson1999photodissociation}
Jackson, J.~M., \& Kraemer, K.~E. 1999, The Astrophysical Journal, 512, 260

\bibitem[{Jackson {et~al.}(2024)Jackson, Whitaker, Chambers, Simon, Guevara, Allingham, Patterson, Killerby-Smith, Askew, Vandenberg, {et~al.}}]{jackson2024absorption}
Jackson, J.~M., Whitaker, J.~S., Chambers, E., {et~al.} 2024, The Astrophysical Journal, 965, 187

\bibitem[{Jeffries(2007)}]{jeffries2007distance}
Jeffries, R. 2007, Monthly Notices of the Royal Astronomical Society, 376, 1109

\bibitem[{Kabanovic {et~al.}(2022)Kabanovic, Schneider, Ossenkopf-Okada, Falasca, G{\"u}sten, Stutzki, Simon, Buchbender, Anderson, Bonne, {et~al.}}]{kabanovic2022self}
Kabanovic, S., Schneider, N., Ossenkopf-Okada, V., {et~al.} 2022, Astronomy \& Astrophysics, 659, A36

\bibitem[{Karim {et~al.}(2023)Karim, Pound, Tielens, Tiwari, Bonne, Wolfire, Schneider, Kavak, Mundy, Simon, {et~al.}}]{karim2023sofia}
Karim, R.~L., Pound, M.~W., Tielens, A.~G., {et~al.} 2023, The Astronomical Journal, 166, 240

\bibitem[{Karr \& Martin(2003)}]{karr2003triggered}
Karr, J., \& Martin, P. 2003, The Astrophysical Journal, 595, 900

\bibitem[{Keto(2002)}]{keto2002evolution}
Keto, E. 2002, The Astrophysical Journal, 580, 980

\bibitem[{Kirsanova {et~al.}(2017)Kirsanova, Sobolev, \& Thomasson}]{kirsanova2017gas}
Kirsanova, M., Sobolev, A., \& Thomasson, M. 2017, arXiv preprint arXiv:1705.02197

\bibitem[{Kirsanova {et~al.}(2020)Kirsanova, Ossenkopf-Okada, Anderson, Boley, Bieging, Pavlyuchenkov, Luisi, Schneider, Andersen, Samal, {et~al.}}]{kirsanova2020pdr}
Kirsanova, M.~S., Ossenkopf-Okada, V., Anderson, L., {et~al.} 2020, Monthly Notices of the Royal Astronomical Society, 497, 2651

\bibitem[{Koo(1997)}]{koo1997hi}
Koo, B.-C. 1997, The Astrophysical Journal Supplement Series, 108, 489

\bibitem[{Krabbe {et~al.}(2008)Krabbe, Mehlert, Krabbe, \& Mehlert}]{Krabbe2008}
Krabbe, A., Mehlert, D., Krabbe, A., \& Mehlert, D. 2008, Naturwissenschaften, 95, 361

\bibitem[{Kuhn {et~al.}(2019)Kuhn, Hillenbrand, Sills, Feigelson, \& Getman}]{kuhn2019kinematics}
Kuhn, M.~A., Hillenbrand, L.~A., Sills, A., Feigelson, E.~D., \& Getman, K.~V. 2019, The Astrophysical Journal, 870, 32

\bibitem[{Lacy {et~al.}(2002)Lacy, Richter, Greathouse, Jaffe, \& Zhu}]{lacy2002texes}
Lacy, J., Richter, M., Greathouse, T., Jaffe, D., \& Zhu, Q. 2002, Publications of the Astronomical Society of the Pacific, 114, 153

\bibitem[{Lancaster {et~al.}(2024)Lancaster, Ostriker, Kim, Kim, \& Bryan}]{lancaster2024geometry}
Lancaster, L., Ostriker, E.~C., Kim, C.-G., Kim, J.-G., \& Bryan, G.~L. 2024, arXiv preprint arXiv:2405.02396

\bibitem[{Lefloch {et~al.}(1997)Lefloch, Lazareff, \& Castets}]{lefloch1997triggered}
Lefloch, B., Lazareff, B., \& Castets, A. 1997, in Herbig-Haro Flows and the Birth of Stars, Vol. 182, 15

\bibitem[{Lenz \& Ayres(1992)}]{lenz1992errors}
Lenz, D.~D., \& Ayres, T.~R. 1992, Publications of the Astronomical Society of the Pacific, 104, 1104

\bibitem[{Lim \& De~Buizer(2019)}]{lim2019surveying}
Lim, W., \& De~Buizer, J.~M. 2019, The Astrophysical Journal, 873, 51

\bibitem[{Lim {et~al.}(2020)Lim, De~Buizer, \& Radomski}]{lim2020surveying}
Lim, W., De~Buizer, J.~M., \& Radomski, J.~T. 2020, The Astrophysical Journal, 888, 98

\bibitem[{Luisi {et~al.}(2021)Luisi, Anderson, Schneider, Simon, Kabanovic, G{\"u}sten, Zavagno, Broos, Buchbender, Guevara, {et~al.}}]{luisi2021stellar}
Luisi, M., Anderson, L.~D., Schneider, N., {et~al.} 2021, Science Advances, 7, eabe9511

\bibitem[{Luong {et~al.}(2011)Luong, Motte, Schuller, Schneider, Bontemps, Schilke, Menten, Heitsch, Wyrowski, Carlhoff, {et~al.}}]{luong2011w43}
Luong, Q.~N., Motte, F., Schuller, F., {et~al.} 2011, Astronomy \& Astrophysics, 529, A41

\bibitem[{Lynds \& Oneil~Jr(1985)}]{lynds1985optical}
Lynds, B.~T., \& Oneil~Jr, E.~J. 1985, Astrophysical Journal, Part 1 (ISSN 0004-637X), vol. 294, July 15, 1985, p. 578-590., 294, 578

\bibitem[{Mart{\'\i}n-Hern{\'a}ndez {et~al.}(2008)Mart{\'\i}n-Hern{\'a}ndez, Peeters, \& Tielens}]{martin2008mid}
Mart{\'\i}n-Hern{\'a}ndez, N., Peeters, E., \& Tielens, A. 2008, Astronomy \& Astrophysics, 489, 1189

\bibitem[{Martins {et~al.}(2010)Martins, Pomar{\`e}s, Deharveng, Zavagno, \& Bouret}]{martins2010near}
Martins, F., Pomar{\`e}s, M., Deharveng, L., Zavagno, A., \& Bouret, J. 2010, Astronomy \& Astrophysics, 510, A32

\bibitem[{Martins {et~al.}(2005)Martins, Schaerer, \& Hillier}]{martins2005new}
Martins, F., Schaerer, D., \& Hillier, D.~J. 2005, Astronomy \& Astrophysics, 436, 1049

\bibitem[{McBreen {et~al.}(1982)McBreen, Fazio, \& Jaffe}]{mcbreen1982high}
McBreen, B., Fazio, G., \& Jaffe, D. 1982, Astrophysical Journal, Part 1, vol. 254, Mar. 1, 1982, p. 126-131., 254, 126

\bibitem[{Moscadelli \& Goddi(2014)}]{moscadelli2014multiple}
Moscadelli, L., \& Goddi, C. 2014, Astronomy \& Astrophysics, 566, A150

\bibitem[{Ochsendorf {et~al.}(2017)Ochsendorf, Zinnecker, Nayak, Bally, Meixner, Jones, Indebetouw, \& Rahman}]{ochsendorf2017star}
Ochsendorf, B.~B., Zinnecker, H., Nayak, O., {et~al.} 2017, Nature Astronomy, 1, 784

\bibitem[{Pabst {et~al.}(2024)Pabst, Goicoechea, Cuadrado, Salas, Tielens, \& Marcelino}]{pabst2024multiline}
Pabst, C., Goicoechea, J., Cuadrado, S., {et~al.} 2024, arXiv preprint arXiv:2404.17963

\bibitem[{Pabst {et~al.}(2020)Pabst, Goicoechea, Teyssier, Bern{\'e}, Higgins, Chambers, Kabanovic, G{\"u}sten, Stutzki, \& Tielens}]{pabst2020expanding}
Pabst, C., Goicoechea, J.~R., Teyssier, D., {et~al.} 2020, Astronomy \& Astrophysics, 639, A2

\bibitem[{Pabst {et~al.}(2021)Pabst, Goicoechea, Hacar, Teyssier, Bern{\'e}, Wolfire, Higgins, Chambers, Kabanovic, G{\"u}sten, {et~al.}}]{pabst2021ii}
Pabst, C., Goicoechea, J., Hacar, A., {et~al.} 2021

\bibitem[{Pabst {et~al.}(2022)Pabst, Goicoechea, Hacar, Teyssier, Bern{\'e}, Wolfire, Higgins, Chambers, Kabanovic, Gusten, {et~al.}}]{pabst2022158}
Pabst, C., Goicoechea, J.~R., Hacar, A., {et~al.} 2022

\bibitem[{Pietrzy{\'n}ski {et~al.}(2019)Pietrzy{\'n}ski, Graczyk, Gallenne, Gieren, Thompson, Pilecki, Karczmarek, G{\'o}rski, Suchomska, Taormina, {et~al.}}]{pietrzynski2019distance}
Pietrzy{\'n}ski, G., Graczyk, D., Gallenne, A., {et~al.} 2019, Nature, 567, 200

\bibitem[{Pomar{\`e}s {et~al.}(2009)Pomar{\`e}s, Zavagno, Deharveng, Cunningham, Jones, Kurtz, Russeil, Caplan, \& Comer{\'o}n}]{pomares2009triggered}
Pomar{\`e}s, M., Zavagno, A., Deharveng, L., {et~al.} 2009, Astronomy \& Astrophysics, 494, 987

\bibitem[{Povich {et~al.}(2007)Povich, Stone, Churchwell, Zweibel, Wolfire, Babler, Indebetouw, Meade, \& Whitney}]{povich2007multiwavelength}
Povich, M.~S., Stone, J.~M., Churchwell, E., {et~al.} 2007, The Astrophysical Journal, 660, 346

\bibitem[{Raga {et~al.}(2012)Raga, Cant{\'o}, \& Rodr{\'\i}guez}]{raga2012universal}
Raga, A.~C., Cant{\'o}, J., \& Rodr{\'\i}guez, L.~F. 2012, Revista mexicana de astronom{\'\i}a y astrof{\'\i}sica, 48, 149

\bibitem[{Ragan {et~al.}(2014)Ragan, Henning, Tackenberg, Beuther, Johnston, Kainulainen, \& Linz}]{ragan2014giant}
Ragan, S.~E., Henning, T., Tackenberg, J., {et~al.} 2014, Astronomy \& Astrophysics, 568, A73

\bibitem[{Reid {et~al.}(2014)Reid, Menten, Brunthaler, Zheng, Dame, Xu, Wu, Zhang, Sanna, Sato, {et~al.}}]{reid2014trigonometric}
Reid, M., Menten, K., Brunthaler, A., {et~al.} 2014, The Astrophysical Journal, 783, 130

\bibitem[{Risacher {et~al.}(2018)Risacher, G{\"u}sten, Stutzki, H{\"u}bers, Aladro, Bell, Buchbender, B{\"u}chel, Csengeri, Duran, {et~al.}}]{risacher2018upgreat}
Risacher, C., G{\"u}sten, R., Stutzki, J., {et~al.} 2018, Journal of Astronomical Instrumentation, 7, 1840014

\bibitem[{Rosen {et~al.}(2014)Rosen, Lopez, Krumholz, \& Ramirez-Ruiz}]{rosen2014gone}
Rosen, A.~L., Lopez, L.~A., Krumholz, M.~R., \& Ramirez-Ruiz, E. 2014, Monthly Notices of the Royal Astronomical Society, 442, 2701

\bibitem[{Roshi {et~al.}(2005)Roshi, Balser, Bania, Goss, \& De~Pree}]{roshi20058}
Roshi, D.~A., Balser, D.~S., Bania, T.~M., Goss, W.~M., \& De~Pree, C. 2005, The Astrophysical Journal, 625, 181

\bibitem[{Russeil {et~al.}(2016)Russeil, Tig{\'e}, Adami, Anderson, Schneider, Zavagno, Samal, Amram, Guennou, Le~Coarer, {et~al.}}]{russeil2016ngc}
Russeil, D., Tig{\'e}, J., Adami, C., {et~al.} 2016, Astronomy \& Astrophysics, 587, A135

\bibitem[{Rygl {et~al.}(2012)Rygl, Brunthaler, Sanna, Menten, Reid, Van~Langevelde, Honma, Torstensson, \& Fujisawa}]{rygl2012parallaxes}
Rygl, K., Brunthaler, A., Sanna, A., {et~al.} 2012, Astronomy \& Astrophysics, 539, A79

\bibitem[{Saha {et~al.}(2024)Saha, Tej, Liu, Liu, Garay, Goldsmith, Lee, He, Juvela, Bronfman, {et~al.}}]{saha2024direct}
Saha, A., Tej, A., Liu, H.-L., {et~al.} 2024, The Astrophysical Journal Letters, 970, L40

\bibitem[{Sandell {et~al.}(2020)Sandell, Wright, G{\"u}sten, Wiesemeyer, Reyes, Mookerjea, \& Corder}]{sandell2020NGC}
Sandell, G., Wright, M., G{\"u}sten, R., {et~al.} 2020, The Astrophysical Journal, 904, 139

\bibitem[{Schneider {et~al.}(2020)Schneider, Simon, Guevara, Buchbender, Higgins, Okada, Stutzki, G{\"u}sten, Anderson, Bally, {et~al.}}]{schneider2020feedback}
Schneider, N., Simon, R., Guevara, C., {et~al.} 2020, Publications of the Astronomical Society of the Pacific, 132, 104301

\bibitem[{Schneider {et~al.}(2023)Schneider, Bonne, Bontemps, Kabanovic, Simon, Ossenkopf-Okada, Buchbender, Stutzki, Mertens, Ricken, {et~al.}}]{schneider2023ionized}
Schneider, N., Bonne, L., Bontemps, S., {et~al.} 2023, Nature Astronomy, 7, 546

\bibitem[{Schneider {et~al.}(2012)Schneider, Csengeri, Hennemann, Motte, Didelon, Federrath, Bontemps, Di~Francesco, Arzoumanian, Minier, {et~al.}}]{schneider2012cluster}
Schneider, N.~e., Csengeri, T., Hennemann, M., {et~al.} 2012, Astronomy \& Astrophysics, 540, L11

\bibitem[{Shimoikura {et~al.}(2018)Shimoikura, Dobashi, Nakamura, Shimajiri, \& Sugitani}]{10.1093/pasj/psy115}
Shimoikura, T., Dobashi, K., Nakamura, F., Shimajiri, Y., \& Sugitani, K. 2018, Publications of the Astronomical Society of Japan, 71, \dodoi{10.1093/pasj/psy115}

\bibitem[{Spitzer~Jr(1968)}]{spitzer1968diffuse}
Spitzer~Jr, L. 1968, Interscience Tracts on Physics and Astronomy, 28

\bibitem[{Stacey {et~al.}(1991)Stacey, Townes, Poglitsch, Madden, Jackson, Herrmann, Genzel, \& Geis}]{stacey1991optical}
Stacey, G., Townes, C., Poglitsch, A., {et~al.} 1991, Astrophysical Journal, Part 2-Letters (ISSN 0004-637X), vol. 382, Nov. 20, 1991, p. L37-L41., 382, L37

\bibitem[{Strömgren(1939)}]{Stromgren1939}
Strömgren, B. 1939, The Astrophysical Journal, 89, 526

\bibitem[{Tielens \& Hollenbach(1985)}]{tielens1985photodissociation}
Tielens, A., \& Hollenbach, D. 1985, Astrophysical Journal, Vol. 291, NO. 2/APR15, P. 747, 1985, 291, 747

\bibitem[{Tielens(2005)}]{Tielens2005}
Tielens, A. G. G.~M. 2005, The Astrophysical Journal

\bibitem[{Tiwari {et~al.}(2019)Tiwari, Menten, Wyrowski, P{\'e}rez-Beaupuits, Lee, \& Kim}]{tiwari2019observational}
Tiwari, M., Menten, K., Wyrowski, F., {et~al.} 2019, Astronomy \& Astrophysics, 626, A28

\bibitem[{Tiwari {et~al.}(2021)Tiwari, Karim, Pound, Wolfire, Jacob, Buchbender, G{\"u}sten, Guevara, Higgins, Kabanovic, {et~al.}}]{tiwari2021sofia}
Tiwari, M., Karim, R., Pound, M., {et~al.} 2021, The Astrophysical Journal, 914, 117

\bibitem[{Tothill {et~al.}(2008)Tothill, Gagn{\'e}, Stecklum, \& Kenworthy}]{tothill2008lagoon}
Tothill, N.~F., Gagn{\'e}, M., Stecklum, B., \& Kenworthy, M.~A. 2008, arXiv preprint arXiv:0809.3380

\bibitem[{Townsley {et~al.}(2018)Townsley, Broos, Garmire, Anderson, Feigelson, Naylor, \& Povich}]{townsley2018massive}
Townsley, L.~K., Broos, P.~S., Garmire, G.~P., {et~al.} 2018, The Astrophysical Journal Supplement Series, 235, 43

\bibitem[{Townsley {et~al.}(2014)Townsley, Broos, Garmire, Bouwman, Povich, Feigelson, Getman, \& Kuhn}]{townsley2014massive}
---. 2014, The Astrophysical Journal Supplement Series, 213, 1

\bibitem[{Townsley {et~al.}(2019)Townsley, Broos, Garmire, \& Povich}]{townsley2019massive}
Townsley, L.~K., Broos, P.~S., Garmire, G.~P., \& Povich, M.~S. 2019, The Astrophysical Journal Supplement Series, 244, 28

\bibitem[{Trevi{\~n}o-Morales {et~al.}(2019)Trevi{\~n}o-Morales, Fuente, S{\'a}nchez-Monge, Kainulainen, Didelon, Suri, Schneider, Ballesteros-Paredes, Lee, Hennebelle, {et~al.}}]{trevino2019dynamics}
Trevi{\~n}o-Morales, S., Fuente, A., S{\'a}nchez-Monge, {\'A}., {et~al.} 2019, Astronomy \& Astrophysics, 629, A81

\bibitem[{Verma {et~al.}(1994)Verma, Bisht, Ghosh, Iyengar, Rengarajan, \& Tandon}]{verma1994far}
Verma, R., Bisht, R., Ghosh, S., {et~al.} 1994, Astronomy and Astrophysics (ISSN 0004-6361), vol. 284, no. 3, p. 936-948, 284, 936

\bibitem[{Walch {et~al.}(2015)Walch, Whitworth, Bisbas, Hubber, \& W{\"u}nsch}]{walch2015comparing}
Walch, S., Whitworth, A., Bisbas, T., Hubber, D., \& W{\"u}nsch, R. 2015, Monthly Notices of the Royal Astronomical Society, 452, 2794

\bibitem[{Weaver {et~al.}(1977)Weaver, McCray, Castor, Shapiro, \& Moore}]{weaver1977interstellar}
Weaver, R., McCray, R., Castor, J., Shapiro, P., \& Moore, R. 1977, Astrophysical Journal, Part 1, vol. 218, Dec. 1, 1977, p. 377-395., 218, 377

\bibitem[{Wenger {et~al.}(2018)Wenger, Balser, Anderson, \& Bania}]{wenger2018kinematic}
Wenger, T.~V., Balser, D.~S., Anderson, L., \& Bania, T. 2018, The Astrophysical Journal, 856, 52

\bibitem[{Werner {et~al.}(1979)Werner, Becklin, Gatley, Matthews, Neugebauer, \& Wynn-Williams}]{werner1979infrared}
Werner, M., Becklin, E., Gatley, I., {et~al.} 1979, Monthly Notices of the Royal Astronomical Society, 188, 463

\bibitem[{Whitney {et~al.}(2004)Whitney, Indebetouw, Babler, Meade, Watson, Wolff, Wolfire, Clemens, Bania, Benjamin, {et~al.}}]{whitney2004glimpse}
Whitney, B., Indebetouw, R., Babler, B., {et~al.} 2004, The Astrophysical Journal Supplement Series, 154, 315

\bibitem[{Wolfire {et~al.}(2022)Wolfire, Vallini, \& Chevance}]{wolfire2022photodissociation}
Wolfire, M.~G., Vallini, L., \& Chevance, M. 2022, Annual Review of Astronomy and Astrophysics, 60, 247

\bibitem[{Xu {et~al.}(2019)Xu, Zavagno, Yu, Liu, Xu, Yuan, Zhang, Zhang, Zhang, Ning, {et~al.}}]{xu2019effects}
Xu, J.-L., Zavagno, A., Yu, N., {et~al.} 2019, Astronomy \& Astrophysics, 627, A27

\bibitem[{Xu {et~al.}(2011)Xu, Moscadelli, Reid, Menten, Zhang, Zheng, \& Brunthaler}]{xu2011trigonometric}
Xu, Y., Moscadelli, L., Reid, M., {et~al.} 2011, The Astrophysical Journal, 733, 25

\bibitem[{Zamora-Avil{\'e}s {et~al.}(2019)Zamora-Avil{\'e}s, V{\'a}zquez-Semadeni, Gonz{\'a}lez, Franco, Shore, Hartmann, Ballesteros-Paredes, Banerjee, \& K{\"o}rtgen}]{zamora2019structure}
Zamora-Avil{\'e}s, M., V{\'a}zquez-Semadeni, E., Gonz{\'a}lez, R.~F., {et~al.} 2019, Monthly Notices of the Royal Astronomical Society, 487, 2200

\bibitem[{Zavagno {et~al.}(2006)Zavagno, Deharveng, Comer{\'o}n, Brand, Massi, Caplan, \& Russeil}]{zavagno2006triggered}
Zavagno, A., Deharveng, L., Comer{\'o}n, F., {et~al.} 2006, Astronomy \& Astrophysics, 446, 171

\bibitem[{Zavagno {et~al.}(2010)Zavagno, Russeil, Motte, Anderson, Deharveng, Rod{\'o}n, Bontemps, Abergel, Baluteau, Sauvage, {et~al.}}]{zavagno2010star}
Zavagno, A., Russeil, D., Motte, F., {et~al.} 2010, Astronomy \& Astrophysics, 518, L81

\bibitem[{Zhu {et~al.}(2005)Zhu, Lacy, Jaffe, Richter, \& Greathouse}]{zhu2005mass}
Zhu, Q.-F., Lacy, J.~H., Jaffe, D.~T., Richter, M.~J., \& Greathouse, T.~K. 2005, The Astrophysical Journal, 631, 381

\end{thebibliography}

\end{document}